\documentclass[journal, twoside]{IEEEtran}

\usepackage{arydshln}

\usepackage[table]{xcolor}

\usepackage{threeparttable}

\usepackage{tikz}

\usetikzlibrary{shapes.misc}
\tikzset{
  every round node/.style={
    draw,
    shape=rounded rectangle,
    rounded rectangle arc length=180,
    inner sep=+.15em,
    text depth=+.1ex},
  light/.style={fill=none, text=none},
  dark/.style={fill=none, text=none}}

\newcommand*\mcircled[5]{\tikz[baseline=(char.base)]{
    \node[shape=circle, fill=#2, draw=#3, text=#4, scale=#5, inner sep=2pt] (char) {#1};}}

\usepackage{multirow}
\usepackage{threeparttable}

\usepackage{adjustbox}

\usepackage{booktabs}

\usepackage{bm}

\usepackage{ifpdf}

\usepackage{xcolor}

\usepackage{listings}

\usepackage{color, soul}

\usepackage{soul}

\usepackage{balance}

\usepackage{float}

\usepackage{enumitem}

\newsavebox{\myimage}

\usepackage[acronym, nomain]{glossaries}
\newacronym[longplural=downlink data]{dld}{DLD}{downlink data}
\newacronym[longplural=uplink data]{uld}{ULD}{uplink data}
\newacronym{adc}{ADC}{analog-to-digital converter}
\newacronym{ad}{A/D}{analog-to-digital}
\newacronym{afe}{AFE}{analog-front end}
\newacronym{agu}{AGU}{address generation unit}
\newacronym{ahb}{AHB}{Advanced High-Performance Bus}
\newacronym{asic}{ASIC}{application-specific integrated circuit}
\newacronym{asip}{ASIP}{application-specific instruction set processor}
\newacronym{awgn}{AWGN}{additive white Gaussian noise}
\newacronym{bb}{BB}{baseband}
\newacronym{ber}{BER}{Bit Error Rate}
\newacronym{bi}{BI}{bus interface}
\newacronym{bs}{BS}{base station}
\newacronym{bw}{BW}{bandwidth}
\newacronym{ce}{CE}{channel estimation}
\newacronym{cm}{CM}{core manager}
\newacronym{cp}{CP}{cyclic prefix}
\newacronym{cpu}{CPU}{central processing unit}
\newacronym{csi}{CSI}{channel state information}
\newacronym{cs}{CS}{circuit-switched}
\newacronym{dac}{DAC}{digital-to-analog converter}
\newacronym{dfe}{DFE}{digital front end}
\newacronym{dft}{DFT}{Discrete Fourier Transform}
\newacronym{dlp}{DLP}{data-level parallelism}
\newacronym{dl}{DL}{down-link}
\newacronym{dma}{DMA}{direct memory access}
\newacronym{dsp}{DSP}{digital signal processing}
\newacronym{dram}{DRAM}{Dynamic \gls{ram}}
\newacronym{dlx}{DLX}{Deluxe}
\newacronym{dvs}{DVS}{dynamic voltage scaling}
\newacronym{ecc}{ECC}{error-correcting code}
\newacronym{eda}{EDA}{electronic design automation}
\newacronym{evm}{EVM}{error vector magnitude}
\newacronym{fdd}{FDD}{frequency-division duplex}
\newacronym{fe}{FE}{Front-End}
\newacronym{fft}{FFT}{fast fourier transform}
\newacronym{fifo}{FIFO}{first-in first-out}
\newacronym{fpga}{FPGA}{Field Programmable Gate Array}
\newacronym{gals}{GALS}{globally-asynchronous locally-synchronous}
\newacronym{ggm}{GGM}{general Gram matrix}
\newacronym{gpp}{GPP}{General Purpose Processor}
\newacronym{gpu}{GPU}{graphical processing unit}
\newacronym{gp}{GP}{general-purpose}
\newacronym{gps}{GPS}{Global Positioning System}
\newacronym{hdl}{HDL}{hardware description language}
\newacronym{hls}{HLS}{high-level synthesis}
\newacronym{hw}{HW}{hardware}
\newacronym{ifft}{IFFT}{inverse \gls{fft}}
\newacronym{iid}{iid}{independent and identically distributed}
\newacronym{ip}{IP}{intellectual property}
\newacronym{iot}{IoT}{internet of things}
\newacronym{isa}{ISA}{instruction set architecture}
\newacronym{ltea}{LTE-A}{long-term evolution advanced}
\newacronym{lte}{LTE}{long-term evolution}
\newacronym{lumami}{LuMaMi}{Lund University \gls{mami}}
\newacronym{lut}{LUT}{lookup-table}
\newacronym{los}{LOS}{line-of-sight}
\newacronym{macl}{MACL}{medium-access control layer}
\newacronym{mac}{MAC}{multiply-accumulate}
\newacronym{mami}{MaMi}{massive \gls{mimo}}
\newacronym{mf}{MF}{matched-filter}
\newacronym{mimo}{MIMO}{multiple-input multiple-output}
\newacronym{mrc}{MRC}{maximum ratio combining}
\newacronym{mrt}{MRT}{maximum ratio transmission}
\newacronym{mumimo}{MU-MIMO}{multi-user \gls{mimo}}
\newacronym{mmse}{MMSE}{minimum mean-square error}
\newacronym{mse}{MSE}{mean-square error}
\newacronym{ni}{NI}{National Instruments}
\newacronym{nlos}{NLOS}{non-line-of-sight}
\newacronym{noc}{NoC}{network-on chip}
\newacronym{ocp}{OCP}{Open Core Protocol}
\newacronym{ofdm}{OFDM}{Orthogonal Frequency Division Multiplexing}
\newacronym{pa}{PA}{power amplifier}
\newacronym{p2p}{P2P}{peer-to-peer}
\newacronym{pap}{PAP}{per-antenna processing}
\newacronym{pcie}{PCIe}{Peripheral Component Interconnect Express}
\newacronym{pe}{PE}{processing element}
\newacronym{per}{PER}{packet error rate}
\newacronym{psp}{PSP}{per-subcarrier processing}
\newacronym{ps}{PS}{packet-switched}
\newacronym{pup}{PUP}{per-user processing}
\newacronym{papr}{PAPR}{peak-to-average power ratio}
\newacronym{pss}{PSS}{Primary Synchronisation Signal}
\newacronym{qam}{QAM}{quadrature amplitude modulation}
\newacronym{qrd}{QRD}{QR-decomposition}
\newacronym{ram}{RAM}{Random Access Memory}
\newacronym{rf}{RF}{radio-frequency}
\newacronym{rlc}{RLC}{Reconfigurable Logic Core}
\newacronym{rgf}{RGF}{register-file}
\newacronym{risc}{RISC}{reduced instruction set computer}
\newacronym{rx}{RX}{receiver}
\newacronym{rzf}{RZF}{regularized \gls{zf}}
\newacronym{sdr}{SDR}{Software-Defined Radio}
\newacronym{sic}{SIC}{successive interference cancelation}
\newacronym{simd}{SIMD}{single instruction multiple data}
\newacronym{snr}{SNR}{signal-to-noise ratio}
\newacronym{soc}{SoC}{system-on chip}
\newacronym{svd}{SVD}{singular-value decomposition}
\newacronym{tdd}{TDD}{time-division duplex}
\newacronym{tdm}{TDM}{time-division multiplexing}
\newacronym{tx}{TX}{transmitter}
\newacronym{ota}{OTA}{over-the-air}
\newacronym{ue}{UE}{user equipment}
\newacronym{Ue}{UE}{User equipment}
\newacronym{upa}{UPA}{user processing accelerator}
\newacronym{ulp}{ULP}{uplink pilot}
\newacronym{ul}{UL}{up-link}
\newacronym{vliw}{VLIW}{very large instruction word}
\newacronym{vlsi}{VLSI}{very-large scale integration}
\newacronym{zf}{ZF}{zero-forcing}
\newacronym{plm}{PLM}{parallel memory}
\newacronym{nop}{NOP}{no operation}
\newacronym{rtl}{RTL}{register transfer level}
\newacronym{alu}{ALU}{arithmetic logic unit}
\newacronym{msb}{MSB}{most significant bit}
\newacronym{lsb}{LSB}{least significant bit}
\newacronym{cnn}{CNN}{convolutional neural network}
\newacronym{blas}{BLAS}{basic linear algebra subprograms}
\newacronym{gemm}{GEMM}{general matrix multiply}
\newacronym{fd-soi}{FD-SOI}{fully depleted silicon-on-insulator}
\newacronym{relu}{ReLU}{rectified linear unit}
\newacronym{gflop}{GLFOP}{giga floating-point operations per second}
\newacronym{gnss}{GNSS}{global navigation satellite system}

\usepackage[caption=false, font=footnotesize]{subfig}

\usepackage{amsmath}

\usepackage{graphicx}
\graphicspath{{./figures/}}

\usepackage[font=small]{caption}
\DeclareCaptionFont{white}{\color{white}}
\DeclareCaptionFormat{listing}{\colorbox[cmyk]{0.43, 0.35, 0.35,0.01}{\parbox{\textwidth}{\hspace{15pt}#1#2#3}}}

\usepackage{algorithm}
\usepackage{algpseudocode}

\usepackage{xspace}

\newcommand*{\captionshift}{-0.25cm}

\presetkeys%
{todonotes}%
{inline,backgroundcolor=yellow}{}
\usepackage[per-mode=symbol]{siunitx}
\usepackage[utf8]{inputenc}
\graphicspath{{./figures/}}
\usepackage{bm}
\usepackage{mathtools}
\usepackage{amssymb}
\usepackage{setspace}
\usepackage{tikz}
\usepackage{threeparttable}
\usetikzlibrary{calc,patterns}
\usetikzlibrary{matrix,positioning}
\usepackage[\ifnum\pdfoutput=1breaklinks\fi]{hyperref}
\usepackage{regexpatch}
\usepackage{subfig}
\usepackage{url}

\usepackage{breakurl}
\usepackage[breaklinks]{hyperref}

\usepackage{listings}

\usepackage{booktabs}
\usepackage{multirow}

\usepackage{svg}

\definecolor{mygreen}{rgb}{0,0.6,0}
\definecolor{mygray}{rgb}{0.5,0.5,0.5}
\definecolor{mBlue}{rgb}{0, 0,0.6}
\definecolor{mymauve}{rgb}{0.58,0,0.82}
\definecolor{backgroundColour}{rgb}{0.98,0.98,0.99}

\lstdefinestyle{CStyle}{
    backgroundcolor=\color{backgroundColour},   
    commentstyle=\color{mGreen},
    keywordstyle=\color{mBlue}\bfseries,
    numberstyle=\tiny\color{mGray},
    stringstyle=\color{mPurple},
    basicstyle=\fontsize{7}{11}\selectfont\ttfamily, 
    breaklines=true,
    framextopmargin=1pt,
    frame=single,
    breakatwhitespace=false,         
    breaklines=true,                 
    captionpos=b, 
    keepspaces=true,                 
    numbers=left,                    
    numbersep=5pt,                  
    showspaces=false,                
    showstringspaces=false,
    showtabs=false,                  
    tabsize=2,
    language=C,
    deletekeywords={...}, 
	morekeywords={*,for, int, bool, false, vcbfloat16,...}, 
    escapeinside={\%*}{*)}, 
	numbersep=0pt, 
	numberstyle=\tiny\color{mygray}, 
}

\usepackage{scalerel}
\usepackage{tikz}
\usetikzlibrary{svg.path}
\definecolor{orcidlogocol}{HTML}{A6CE39}
\tikzset{
    orcidlogo/.pic={
            \fill[orcidlogocol] svg{M256,128c0,70.7-57.3,128-128,128C57.3,256,0,198.7,0,128C0,57.3,57.3,0,128,0C198.7,0,256,57.3,256,128z};
            \fill[white] svg{M86.3,186.2H70.9V79.1h15.4v48.4V186.2z}
            svg{M108.9,79.1h41.6c39.6,0,57,28.3,57,53.6c0,27.5-21.5,53.6-56.8,53.6h-41.8V79.1z M124.3,172.4h24.5c34.9,0,42.9-26.5,42.9-39.7c0-21.5-13.7-39.7-43.7-39.7h-23.7V172.4z}
            svg{M88.7,56.8c0,5.5-4.5,10.1-10.1,10.1c-5.6,0-10.1-4.6-10.1-10.1c0-5.6,4.5-10.1,10.1-10.1C84.2,46.7,88.7,51.3,88.7,56.8z};
    }
}
\newcommand\orcidicon[1]{\href{https://orcid.org/#1}{\mbox{\scalerel*{
\begin{tikzpicture}[yscale=-1,transform shape]
\pic{orcidlogo};
\end{tikzpicture}
}{|}}}}
\usepackage{hyperref} 

\ifpdf

\else

\fi

\begin{document}

	\title{Accelerator-assisted Floating-point ASIP for Communication and Positioning in Massive MIMO Systems}

	\author{
		Mohammad~Attari\orcidicon{0000-0002-5192-4890},~\IEEEmembership{Student~Member,~IEEE,}
		Ove~Edfors\orcidicon{0000-0001-5966-8468},~\IEEEmembership{Senior Member,~IEEE,}, and Liang~Liu\orcidicon{0000-0001-9491-8821},~\IEEEmembership{Member,~IEEE}		
		
		\thanks{M. Attari, O. Edfors, and L. Liu are with the Department of Electrical and Information Technology, Lund University, Box 118, SE-221 00 Lund, Sweden, Email: \{firstname.lastname\}@eit.lth.se}%
	}

	\maketitle

	\begin{abstract}
		This paper presents an implementation of a floating-point-capable \gls{asip} for both communication and positioning tasks using the massive \gls{mimo} technology. The \gls{asip} is geared with vector processing capabilities in the form of \gls{simd}. A dual-pronged accelerator composition assists the processor to tame the heavier mathematical workloads. A standalone systolic array accelerator accompanies the processor to aid with matrix multiplications. A parallel vector memory subsystem provides functionalities to both the processor and the systolic array. Additionally, A \gls{cnn} module accelerator, which is paired with its own separate vector memory, works hand in glove with the processor to take on the positioning task. The processor is synthesized in \SI{22}{\nano\meter} \gls{fd-soi} technology running at a clock frequency of 800~\si{\mega\hertz}. The system achieves a maximum detection throughput of 2.1 Gb/s in a 128$\times$16 massive \gls{mimo} system for the \gls{ue} speed of 50\si{\kilo\meter\per\hour}. The localization throughput settles at around 390 positionings/s.
	\end{abstract}
	

	\IEEEpeerreviewmaketitle

	\glsresetall
	
	\section{Introduction}
	\label{sec:intro}
	\IEEEPARstart{T}{he} current and next generation of wireless networks (5G and beyond) play an integral part in people's everyday lives. These modern wireless communication systems are often combined with positioning systems, where position information is used for various tasks, such as navigation. Positioning capabilities are central to many existing and new applications, while precise positioning is often limited to locations where services such as \gls{gnss} signals are available. Being able to extract position information from signals already available for communication can improve positioning accuracy where signals dedicated for positioning are not available. By extracting information about the propagation environment, in the form of channel estimates, positioning can be performed using wireless communication signals -- a task often addressed by applying machine learning techniques, as they require no calibration at the antenna array.
 
    The aforementioned systems are highly complex and require specialized computing hardware in order to meet their computational demands. Accelerators today are championed as the way forward \cite{10.1145/3282307}, but fixed-function implementations do not always satisfy the ever-expanding requirements and suffer from obsolescence in this volatile field, hence retaining some flexibility scores high on the list of supported features \cite{8416771}. To meet these demands, silicon-efficient design of hardware  goes hand-in-hand with bespoke algorithm design (algorithm-hardware codesign). One of the mechanisms through which this process is facilitated is by using an \gls{asip}. These processors provide some of the much-vaunted programmability of a general-purpose system while enabling \gls{asic}-like performance \cite{9319703, 7446051}. The flip side of this attained flexibility is that some performance is sacrificed, but the advantages outweigh the drawbacks.

    Efficient acceleration of massive \gls{mimo} baseband processing has been subject to prior and ongoing research. The work in \cite{9492455} presented a 128 $\times$ 8 massive \gls{mimo} processing system employing an 8-lane complex \gls{asip}, but did not put an emphasis on matrix computation acceleration. A flexible and salable reconfigurable architecture for baseband massive \gls{mimo} detection was provided in \cite{8911207}, foregoing programmability in exchange for more performance. The authors in \cite{7870260} proposed a fixed-function $128\times8$ massive \gls{mimo} precoder-detector. A non-programmable $128\times16$ massive \gls{mimo} detector utilizing a systolic array architecture was presented in \cite{8310265}. Finally, a \gls{soc} comprised of 8 RISC-V cores was proposed in \cite{9911311}, which harnessed floating-point numbers to accelerate massive \gls{mimo} non-linear algorithms. Non of the these works combine positioning and communication processing in one system.
    
    As this work's contribution, we utilize RISC-V \cite{Waterman:EECS-2016-118} as the base \gls{isa} and augment it with reduced-precision floating-point capabilities. RISC-V is picked as it is an open standard \gls{isa}, with modular design and explicit support for custom extensions, and is predicted to become a major player in the global processor market \cite{10049118}. The work presented in \cite{9801536} is re-designed and, on top of a more streamlined vector core, now benefits from a bigger, more self-sufficient systolic array with floating-point support for complex matrices. Furthermore, the \gls{cnn} accelerator in \cite{9401528} has been incorporated into the design to reinforce the processor's capabilities at handling the positioning process.

    The remainder of this article is organized as follows. Section \ref{sec:communication_and_positioning} briefly lays out the communication and positioning processes, as well as their attendant algorithms. Section \ref{sec:asip_cores_accelerators} switches over to the processor proper and its micro-architecture, while providing a more detailed account of the accelerators attached to the processor. In section \ref{sec:evaluation_sec} evaluation and implementation results are discussed. And finally, section \ref{sec:conclusion} wraps up the article with concluding remarks.
    
	\section{Communication and positioning using Massive MIMO}
	\label{sec:communication_and_positioning}
    \begin{figure}[!t]
		\centering
		\includegraphics[trim=0cm 0cm 0cm 0cm, clip=true, width=.7\columnwidth]{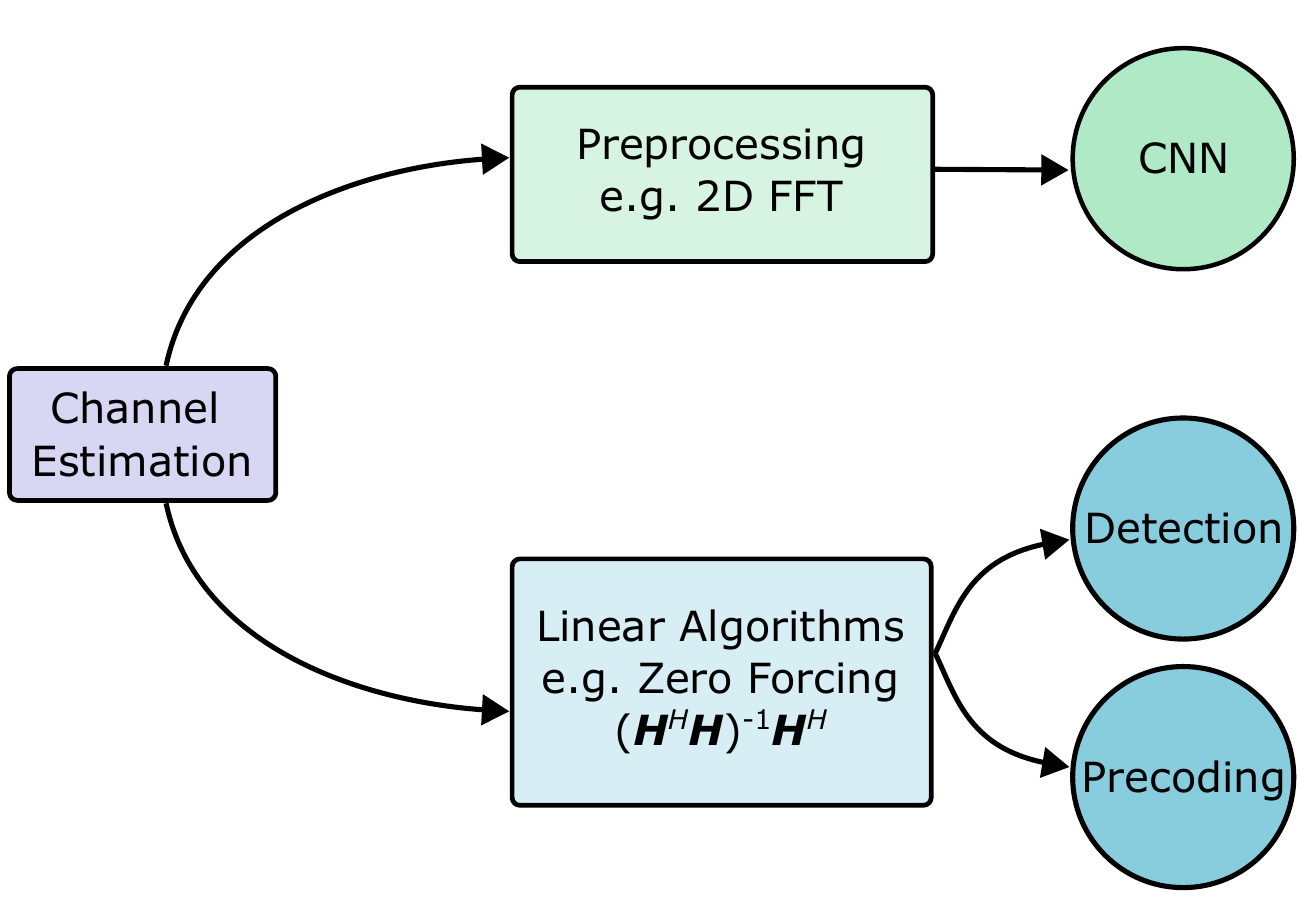}
		\vspace{\captionshift}
		\captionsetup{font={small}}
		\caption{Task mapping.}
		\label{fig:mapping}
	\end{figure}
 
    \subsection{Algorithms and Tasks}
    The communication and positioning tasks both utilize the information acquired through the channel estimation process. Fig. \ref{fig:mapping} diagrams how the channel state information is exploited to enable both positioning and communication. For positioning, the relevant chunk of the channel state information is fed to a pre-processor first, e.g. \gls{fft} \cite{8292280} or spatial covariance matrix and channel impulse response extractor \cite{10330061}, and then the data is forwarded to a \gls{cnn} module for final positioning. In case of \gls{fft}, the \gls{csi} is transformed from the antenna-frequency domain into the angular-delay domain, which creates a sparse $\textit{image}$ for easier digestion in the \gls{cnn}. This is simply carried out as a two-dimensional (2D) \gls{fft} that can be performed as left and right matrix multiplications to carry out the 2D \gls{fft}. The matrix multiplication operation features prominently in the communications algorithms as well, and hence can take advantage of the might of the systolic array. After 2D \gls{fft}, the processed data is steered towards the \gls{cnn}, which has its dedicated module as an accelerator to supercharge the positioning application.
    
    The communication task, on the other hand, takes its relevant part of the channel state and runs a linear algorithm, such as \gls{zf} or \gls{mmse}, to perform either precoding or postcoding (detection). The communication algorithm for a system with $M$ antennas and $K$ \glspl{ue} can be summarized as in the following. An estimate for the transmitted symbol vector $\bm{\hat{y}}$ is acquired by multiplying the $M \times 1$ received data vector $\bm{r}$ by the $K \times M$ detection matrix $\bm{W_\text{det}}$, expressed as
    
	\begin{align}
		\label{eq:uplink_detection}
		\bm{\hat{y}} = \bm{W}_\text{det}\bm{r}
	\end{align}

    The detection matrix in (\ref{eq:uplink_detection}) is obtained from the information contained in the estimated \gls{ul} channel matrix $\bm{\hat{H}}$. For the \gls{zf} algorithm, which is used for analysis in this paper, the $\bm{W_\text{det}}$ is equivalent to the pseudo-inverse of the estimated channel matrix $\bm{H}^{\dagger}$, given as

	\begin{align}
		\label{eq:detection_matrix}
		\bm{H}^{\dagger}=(\bm{\hat{H}}^\mathsf{H}\bm{\hat{H}})^{-1}\bm{\hat{H}}^\mathsf{H}=\bm{W}_\text{det}
	\end{align}       

    \begin{figure}[!t]
		\centering
		\includegraphics[trim=0cm 0cm 0cm 0cm, clip=true, width=.8\columnwidth]{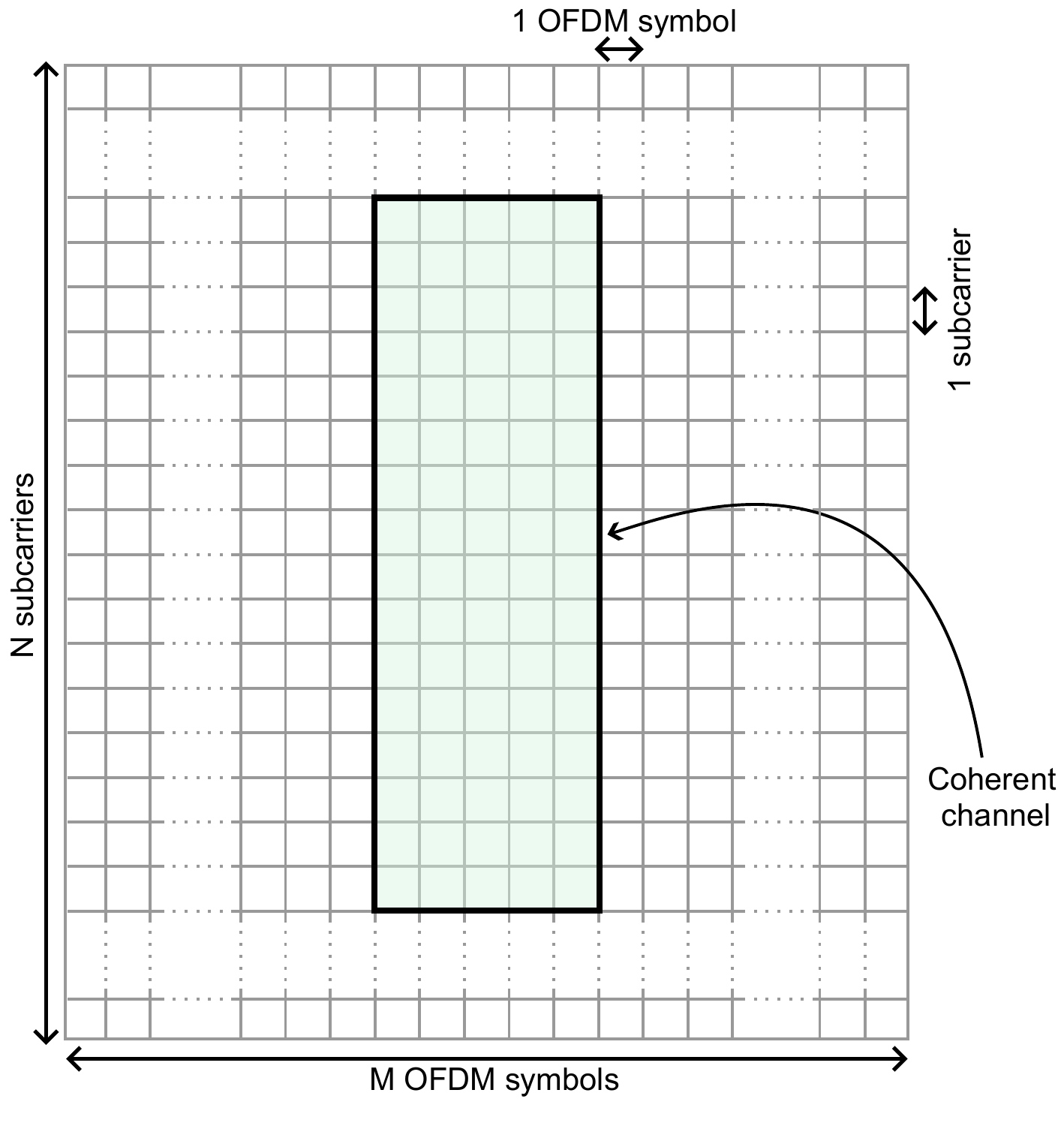}
		\vspace{\captionshift}
		\captionsetup{font={small}}
		\caption{Resource grid.}
		\label{fig:resource_grid}
	\end{figure}

    \begin{table}[!t]
		\centering
		\footnotesize
        \captionsetup{justification=centering}
		\caption{\label{tab:inversions_per_second} Inversions per second.}	
        \renewcommand{\arraystretch}{1.0}
		\begin{tabular}{ c|c|c|c|c|c }
			\hline
            \rotatebox[origin=c]{90}{\textbf{Subcarrier spacing}} &
			\rotatebox[origin=c]{90}{\textbf{Bandwidth}} & \rotatebox[origin=c]{90}{\textbf{~Total subcarriers (N)~}} & \rotatebox[origin=c]{90}{\parbox{2.2cm}{\textbf{\#OFDM symbols}\\ \textbf{~in one second (M)}}} & \rotatebox[origin=c]{90}{\parbox{2.2cm}{\textbf{\#Resource elements}\\ \textbf{~in one second}}} & \rotatebox[origin=c]{90}{\textbf{\#Inversions/s}}\\
			\hline
			\hline
            15 kHz & 20 MHz & 1200 & 14000 & 16.8 M & 210 k\\
            \hline 
            15 kHz & 50 MHz & 3300 & 14000 & 46.2 M & 580 k\\
            \hline 
            60 kHz & 100 MHz & 1650 & 56000 & 92.4 M & 1.16 M\\
			\hline
            60 kHz & 200 MHz & 3300 & 56000 & 184.8 M & 2.3 M\\
			\hline
		\end{tabular}	  
	\end{table}

    \subsection{Computational Analysis}
    As can be observed, the calculation involves matrix and vector manipulations, implying reliance on certain algebraic operations discussed in the next subsection. But here we look at the system from a higher abstraction level in terms of its computational demands. Fig. \ref{fig:resource_grid} lays out the resource grid structure, for a one-second interval, used in the newer generations of communication standards (4G onwards). The structure divides the time-frequency domain into $M$ \gls{ofdm} symbols in one second and $N$ subcarriers. By inspecting this grid we can expose the computational requirements that are put on processing systems. Table \ref{tab:inversions_per_second} lists the requirements for a couple of scenarios based on subcarrier spacing and channel bandwidth. By assuming channel coherency\footnote{Meaning a channel with an unvarying channel impulse response.} for 16 subcarriers and 5 \gls{ofdm} symbols, as depicted by the highlighted box in Fig. \ref{fig:resource_grid}, 210k channel inversions are required every second for a 15 kHz subcarrier spacing and 20 MHz bandwidth. This metric is increased to 2.3M required inversions preformed in one second if we bump up the spacing to 60 kHz and increase the bandwidth to 200 MHz. This should make it clear that the communications processing is putting up stiff challenges to overcome in terms of processing throughput.

    \begin{table}[!t]
		\centering
		\footnotesize
        \captionsetup{justification=centering}
		\caption{\label{tab:needed_operations} Example kernel operations.}	
        \renewcommand{\arraystretch}{1.3}
		\begin{tabular}{ l|l|l  }
			\hline
			\textbf{Description} & \textbf{Symbol} & \textbf{Operation}\\
			\hline
			\hline
            Matrix multiplication & $\bm{A}\times\bm{B}$ & $\bm{C}_{ij}=\sum_k \bm{A}_{ik}\times\bm{B}_{kj}$\\
            \hline
			Hadamard (element-wise) & $\bm{a}\odot\bm{b}$ & $(\bm{a})_i\times{(\bm{b})_i}$\\
			vector product & &\\   
			\hline
            Element-wise vector & $\bm{a}\pm\bm{b}$ & $(\bm{a})_i\pm{(\bm{b})_i}$\\
            addition/subtraction & &\\      
            \hline
            Vector dot product$^*$ & $\bm{a}\cdot\bm{b}$ & $\sum_i (\bm{a}_i\times\overline{\bm{b}_i})$\\
            \hline
            Vector-scalar product & $\bm{a}\times s$ & $(\bm{a})_i\times{s}$\\   
            \hline
            Vector element modify & $\bm{a}[n]=s$ & $\bm{a}_n=s$\\
            \hline
            Vector-vector element & $\bm{a}\times \bm{b}[n]$ & $(\bm{a})_i\times{\bm{b}_n}$\\
            product & & \\
            \hline
			Vector norm & $\lVert\bm{a}\rVert$ & $\sqrt{ \bm{a}\cdot\bm{a}}$\\
			\hline
		\end{tabular}
		\begin{tablenotes}
		    \item * $\overline{b}$ signifies complex conjugation of $b$
		\end{tablenotes}	  
	\end{table}
 
    \subsection{Algebraic Operations}
    Table \ref{tab:needed_operations} lists some of the kernel algebraic operations that need to be supported in order to implement the communications algorithms. The efficient execution of these operations is important, and the processor must provide the necessary facilities to speed them up. Section \ref{sec:asip_cores_accelerators} digs deeper into the processor's micro-architecture and goes into more detail as to how this is accomplished.
    
	\subsection{Data Type}
	In the world of digital processing for embedded systems the fixed-point data type is king. There are usually good reasons behind this:
    
    \begin{itemize}
        \item size: the fixed-point hardware is less complex and hence is smaller. It also needs less memory space
        \item power consumption: smaller circuits lead to less power draw
        \item speed: the simpler design means the critical path is shorter and calculations can be done faster
        \item cost effectiveness: fixed-point hardware can result in cost savings compared to its floating-point counterpart
    \end{itemize}

     \begin{figure*}[t!]
		\centering
		\includegraphics[trim=0cm 0cm 0cm 0cm, clip=true, width=1\textwidth]{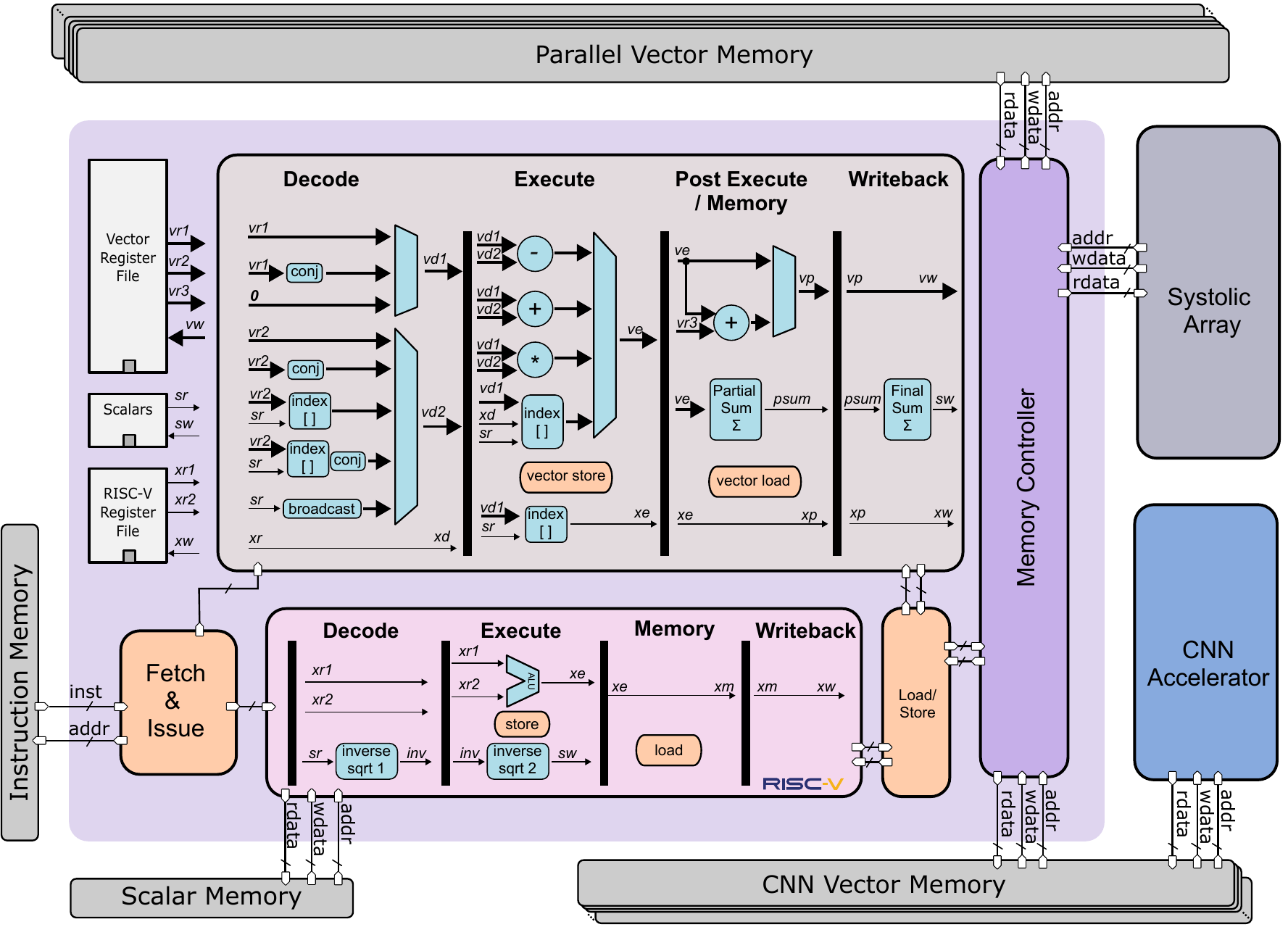}
		\vspace{\captionshift}
		\captionsetup{font={small}}
		\caption{Bird's-eye view of the innards of the processor, illustrating the stylized structure of the RISC-V processor and the vector core surrounded with the different memories and the two accelerators.}
		\label{fig:birds_eye}
	\end{figure*} 
 
    But fixed-point numbers are not the only method of benefiting from real numbers in hardware. On the other side of the spectrum for real numbers we have the floating point number formats. For instance, there are the IEEE standard 32-bit floating point type (float32) and its half-precision sibling (float16), which provide a lot better flexibility and precision at the cost of requiring more resources. These are mostly deployed in systems that have less stringent hardware resource requirements. 
 
    Contrary to popular belief, reduced-precision floating-point formats can be serious contenders to fixed-point implementations \cite{5995137, 9911311}. A rather recent addition to the potpourri of floating-point number representations is the brain floating-point format (bfloat16), which has gained popularity in machine learning training and inferencing tasks, with support from major architectures and accelerators \cite{intel_amx, google_bfloat16_tpu, cerebras_bfloat16}. The bfloat16 data type is a truncated version of the float32 type, occupying 16 bits. It has almost the same dynamic range due to the same number of bits dedicated to the exponent (8 bits), but has only 7 bits in the fractional part, which means it has less precision. 
 
    One of the benefits of the bfloat16 data type is its relaxed requirements on the hardware resources. With a smaller number of bits dedicated to the mantissa (7 bits), the bfloat16 multipliers take up about half the silicon area compared to float16 multipliers, and this is even more pronounced in juxtaposition with float32 (eight times smaller) \cite{google_bfloat16}. Another advantage of the bfloat16 data type is the extremely easy up/down conversion to/from the float32 type. This is simply accomplished by either stripping away the lower 16 bits of the float32 to down-convert, or by appending 16 zeros to the lower bits of the bfloat16 to up-convert.

    In terms of performance, it has been demonstrated that bfloat16 can closely compete with the double-precision floating-point representation sans re-calibration in a massive \gls{mimo} setting \cite{9911311}. While the usage of fixed-point numbers in all stages of computation can lead to a performance loss of 0.5 dB at 0.1\% \gls{ber}, doing so entails extensive simulation and calibration effort to combat the loss. Moreover, the places where massive \gls{mimo} is deployed take on a multitude of scenarios, depending on the number of antennas, number of users, and a mix of \gls{los} and \gls{nlos} propagation \cite{10405907}. This variety further widens the performance gap between fixed-point and floating-point representations, when the number of users is higher, and support for both \gls{los} and \gls{nlos} is needed. The dynamic range granted by floating-point numbers is a boon in dealing with the varied nature of these systems.
    
    For the above-mentioned reasons, bfloat16 has been chosen as the main representation for the system. Specifically, the vector core and the systolic array utilize this data type, while the \gls{cnn} accelerator retains its native fixed-point implementation.

	\subsection{System Parameters}
    Massive \gls{mimo} systems come in a variety of setups in terms of the number of users and the antenna elements. This design adds further flexibility in its support for systems with more antennas and more number of users.
    
	\section{The ASIP Architecture}
	\label{sec:asip_cores_accelerators}
	\subsection{High-level Overview of the Processor}
    The efficient execution of the aforementioned communication and positioning tasks demands special consideration when it comes to the design of the processor. Toward this end, the work presented in \cite{9801536} has been redesigned from scratch and has joined forces with the \gls{cnn} accelerator in \cite{9401528} to tackle both tasks efficiently. The processor now consists of a quartet of principal datapath components, in the form of a scrawny RISC-V scalar core, a brawny 16-lane vector core, a beefed-up, tightly-coupled 16 $\times$ 16 systolic array accelerator, and a \gls{cnn} module accelerator. Fig. \ref{fig:birds_eye} (not to scale) portrays the high-level view of the processor subsystem.
 
	General purpose and light-weight processing is carried out in the scalar core, while the vector core presides over data-parallel workloads in \gls{simd} fashion. The \gls{simd} processing is chosen as it excels at amortizing the overhead incurred by instruction fetch and decode over a larger datapath width. Additionally, a systolic array accelerator works as an offload engine which is tailor-made for computations that are heavy on regular matrix math. Finally, a standalone \gls{cnn} engine takes on the role of expediting the positioning task. 

    \begin{figure}[!t]
		\centering
		\includegraphics[trim=0cm 0cm 0cm 0cm, clip=true, width=.8\columnwidth]{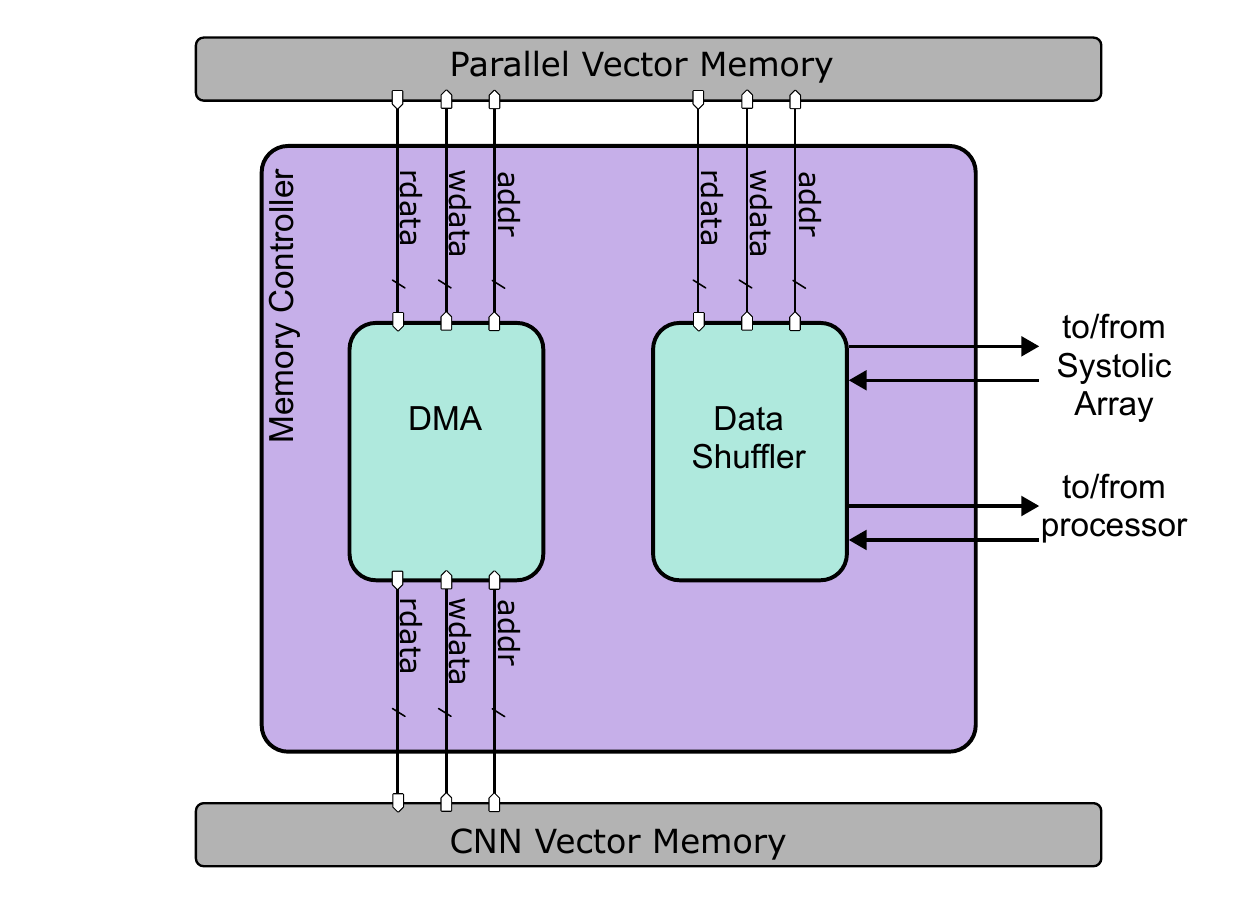}
		\vspace{\captionshift}
		\captionsetup{font={small}}
		\caption{Memory controller.}
		\label{fig:memory_controller}
	\end{figure}
 
    \subsection{Memory System}
    In order to sustain the required throughput, it is necessary to execute communication and positioning simultaneously. In view of this problem, the proposed system exploits two different vector memories. One is shared between the processor and the systolic array, while the other is solely accessible by the \gls{cnn} accelerator. This separation of memories means that the system ends up requiring more memory space, but on the other hand it makes it possible to run communication and positioning tasks in parallel. 

    The shared vector memory is an instance of parallel memory. For readers who wish to take their understanding further, the technical details of this memory are described in \cite{Yang:2017, 8445578}, with the implementation explained fully in \cite{9801536}. Since the systolic array and the vector core access the same vector memory, only one can be actively using it. As a result, while the systolic array is operational, the processor suspends operations until the array is finished, and only then it will resume its normal execution.
 
    One caveat of the two-memory system is that the channel data needs to be copied from the parallel vector memory to the \gls{cnn} vector memory before positioning can commence. This is carried out by a dedicated module inside the memory controller that reads the complex-valued matrix channel data from parallel vector memory, and, after splitting the real and imaginary parts, stores the results as two matrices in the \gls{cnn} vector memory. This is prosecuted by the \gls{dma} module in the memory controller, which is launched by the processor and then performs the copy-and-split action in an unsupervised fashion. 
    
    Fig. \ref{fig:memory_controller} sketches out the block diagram of the memory controller. The data shuffler performs address generation and element swizzling for the parallel vector memory, which enables fast access to the matrix rows/columns. To tap into this feature, the matrices of different dimensions need to be placed in the parallel vector memory in a special manner. The following subsection is devoted to this topic.

	\begin{figure}[!t]
		\centering
		\includegraphics[trim=0cm 0cm 0cm 0cm, clip=true, width=.8\columnwidth]{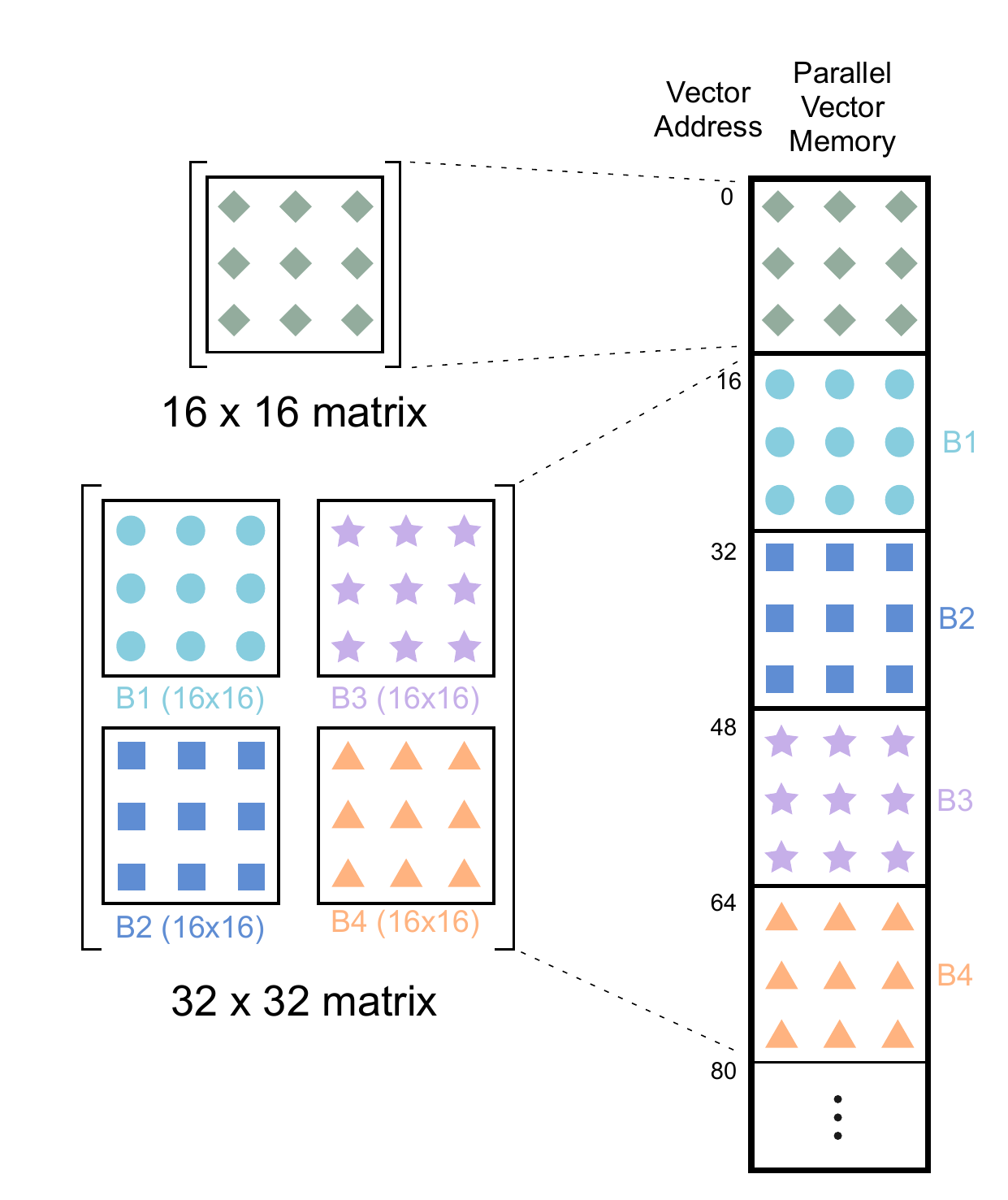}
		\vspace{\captionshift}
		\captionsetup{font={small}}
		\caption{Matrix layout in the parallel vector memory for a 16 $\times$ 16 matrix followed by a 32 $\times$ 32 matrix.}
		\label{fig:matrix_memory_layout}
	\end{figure}
 
	\subsection{Matrix Memory Layout}
    \label{sec:matrix_mem_layout} 
    The matrices are stored in the parallel vector memory as one-dimensional arrays of 16-element vectors. For instance, a 32 $\times$ 32 matrix will be stored as a flat 64-element array of vectors. This matrix flattening is accomplished by first partitioning the matrix into blocks of 16 $\times$ 16 sub-matrices. Secondly, these sub-matrix blocks are placed in a block-wise column-major fashion in the memory. 
    
    Fig. \ref{fig:matrix_memory_layout} depicts how two different matrices are stored in the vector memory. The 16 $\times$ 16 matrix is simply stored as an array with a 16-element-wise access granularity. The matrix of the order 32 $\times$ 32 is laid out in the vector memory as four blocks. The four 16 $\times$ 16 blocks composing the matrix (that is, B1, B2, B3, B4) dwell in the vector memory one after the other. As can be seen, the whole matrix spans over 64 vectors in the memory.

    This way of matrix layout makes it so that access to any column or row of a high-dimensional matrix can be done easily and efficiently. For the 32 $\times$ 32 example above each full row or column access requires two cycles.
    
    \subsection{RISC-V Core}
    The control tower of the processor is materialized as the RISC-V core. It is based on the standard 5-stage pipeline, augmented with extensions for processing bfloat16 numbers, such as the handy inverse square root instruction. The square root unit's implementation is based on the harmonized parabolic synthesis method \cite{8839449}, with a very small footprint, low latency, and good error properties. This instruction's execution path is divided in two stages to break the critical path. 
 
    \subsection{Vector Core}
    Table \ref{tab:needed_operations} summarized some of the kernel operations that require special attention from the processor. The scalar core by itself can not crunch numbers in an aggressive manner to satisfy this. Therefore, to extend the envelope of the processor's capabilities, a vector core is included in the design, which can be thought of as the \gls{simd} heart of the processor. To simplify the design it has the same number of pipeline stages as the baseline RISC-V core (Fig. \ref{fig:birds_eye}). At every pipeline stage some form of processing is carried out, as described in the following: 
    \begin{enumerate}[label=(\alph*)]
        \item \textbf{Decode}: At the decode stage the vectors undergo pre-processing. At this stage the first vector operand supplied to the next stage can be the first vector read from the vector register file, its conjugate, or a zero vector. The second operand has more possibilities. These include the second register file vector passing through intact, indexed into by a scalar value and broadcasted, indexed and conjugated and broadcasted, or a scalar value being broadcasted. Alternatively a scalar value can be broadcast to the next stage. Furthermore, the RISC-V register file can supply a register to be used in vector indexing operations in the following stages.
        \item \textbf{Execute}: The execution stage performs the main operations on the vectors such as addition, subtraction, and multiplication. Vector indexing also happens here. The indexing operation can be used to either read a single value from the first vector into one of the RISC-V registers, or to write a value from the scalar registers into a single element of the first vector, indexed by a RISC-V register.
        \item \textbf{Post Execute}: The \gls{mac} and dot product operations are two of the most useful calculations in linear algebra. This stage helps in implementing these in an efficient manner. The \gls{mac} unit here gives the possibility to add a third vector read from the vector register file to the running sum. As for the dot product, since it needs to take the sum over 16 vector elements, this could create a long critical path and, therefore, it has been implemented as a binary tree adder and broken into two parts, with the first part completed in this stage and the second part done in the writeback stage.
        \item \textbf{Write Back}: The vector result is written back to the vector register file in this stage. The dot product is finalized here and written back to the scalar register file. The vector indexing read operation's result will be written back to the RISC-V register file.
    \end{enumerate}

    \begin{figure}[!t]
		\centering
		\includegraphics[trim=0cm 0cm 0cm 0cm, clip=true, width=1\columnwidth]{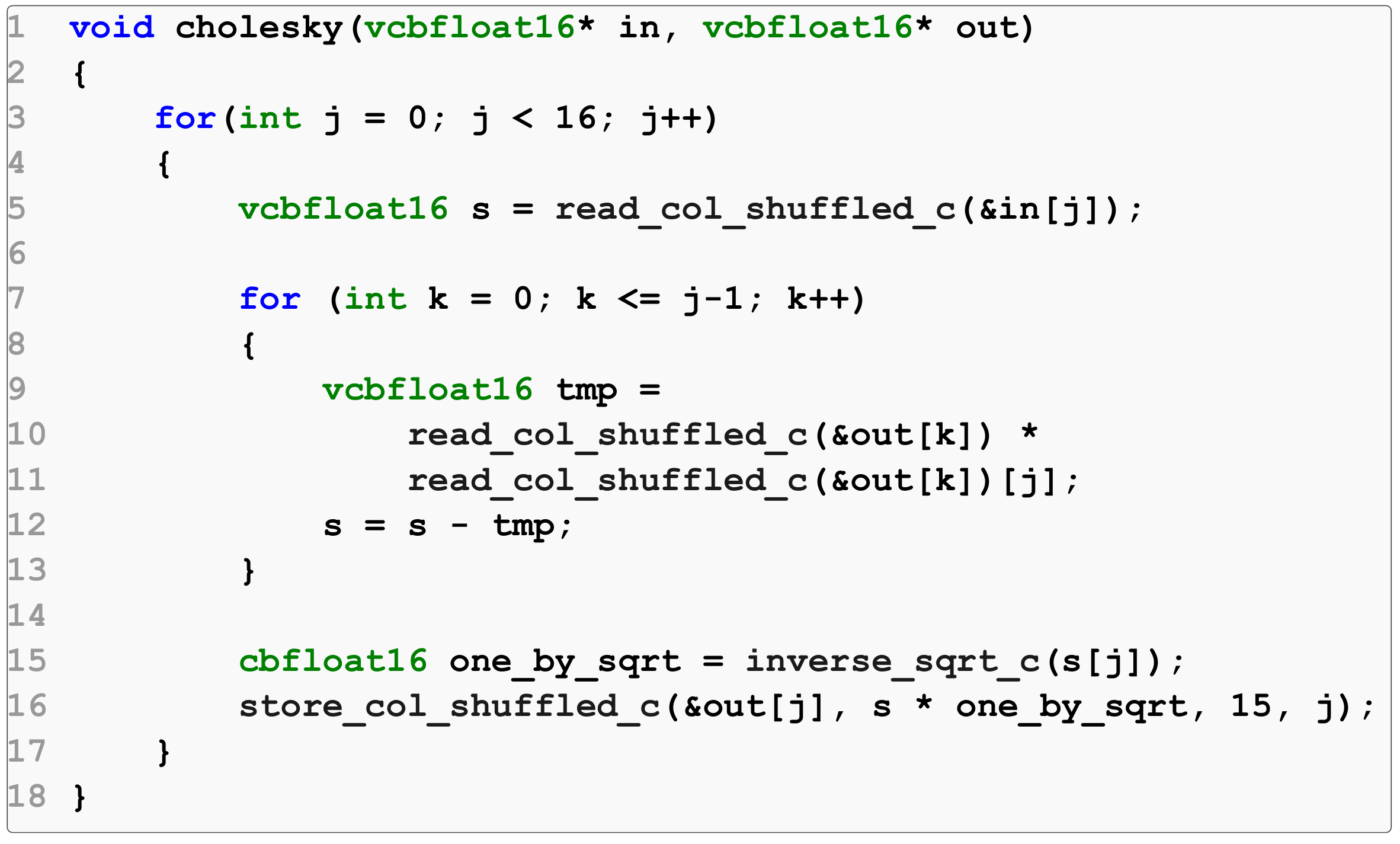}
		\vspace{\captionshift}
		\captionsetup{font={small}}
		\caption{C code that implements the Cholesky algorithm on the \gls{asip}.}
		\label{fig:code_cholesky}
	\end{figure}

    \begin{figure}[!t]
		\centering
		\includegraphics[trim=0cm 0cm 0cm 0cm, clip=true, width=1\columnwidth]{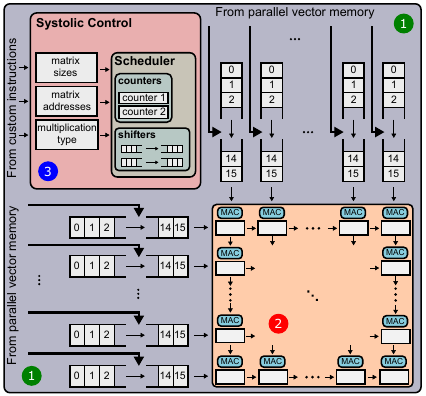}
		\vspace{\captionshift}
		\captionsetup{font={small}}
		\caption{Systolic array accelerator.}
		\label{fig:systolic_array}
	\end{figure}
 
    Fig. \ref{fig:code_cholesky} demonstrates how the processor can perform the Cholesky decomposition by exploiting the hardware-expressive features available through the use of compiler intrinsics. The code is written for a 16 $\times$ 16 input matrix and uses memory-specific and vector-related instructions to carry out the task at hand. The following subsection switches to the topic of acceleration and outlines how it ties to the \gls{asip} and, additionally, shows how the programmer can take advantage of this mechanism while being shielded from the intricacies of the hardware. 
    
    \subsection{Systolic Array}	
    In the communications domain, and in science and engineering disciplines in general, linear algebra plays a significant role. The set of low-level routines that provide the building blocks of common linear algebra operations are categorized as \gls{blas}. There are three levels to these routines with level 3 signifying functions that perform matrix-matrix operations \cite{bb5afff9714941c681fc0e5270168a71}. 
 
    Systolic arrays are suited to the \gls{blas}-3 operations and specifically excel at \gls{gemm}. As such, they have served as souped-up matrix multipliers in computer architecture designs since their inception \cite{kung1979systolic, kung1982systolic}. A key hardware ingredient in achieving high performance is parallelization. Consequently, systolic arrays gang many \glspl{pe} together in a fabric that can work in parallel in a highly regular fashion, but their hardware realizations can take on a multitude of forms \cite{doi:10.1080/00207169808804678}. 

	\begin{table}[!t]
		\centering
		\footnotesize
        \captionsetup{justification=centering}
		\caption{Systolic array vs. the 16-lane vector core for matrix multiplications of varying dimensions (in cycles)}
		\noindent\begin{tabular}{lccc}
			\toprule
			\textbf{Matrix multiplication} & \textbf{Vector} & \textbf{Systolic} & \textbf{Speedup}\\
			\textbf{order} & \textbf{core} & \textbf{array} & \textbf{}\\   
			\midrule
			$(\bm{A}_{16 \times 16})$~*~$(\bm{B}_{16\times 16})$ & 946 & 73 & 12.9$\times$\\
            $(\bm{A}_{32 \times 32})$~*~$(\bm{B}_{32\times 32})$ & 6958 & 400 & 17.4$\times$\\
            $(\bm{A}_{64 \times 64})$~*~$(\bm{B}_{64\times 64})$ & 54k & 2632 & 20.5$\times$\\    
            $(\bm{A}_{128 \times 128})$~*~$(\bm{B}_{128\times 128})$ & 424k & 19k & 22.3$\times$\\  
            $(\bm{A}_{256 \times 256})$~*~$(\bm{B}_{256\times 256})$ & 3367k & 143k & 23.5$\times$\\                
			$(\bm{A}_{16 \times 128})$~*~$(\bm{B}_{128\times 16})$ & 6658 & 304 & 22$\times$\\		
			$(\bm{A}_{16 \times 256})$~*~$(\bm{B}_{256\times 16})$ & 13k & 568 & 22.9$\times$\\
			$(\bm{H}_{16 \times 128})$~*~$(\bm{H}^H_{128\times 16})^{a}$ & 2162 & 168 & 12.9$\times$\\   
			\bottomrule
		\end{tabular}
		\label{tab:systolic_savings}
		\begin{tablenotes}
		    \item  $^{a}$ For the Gramian matrix $(\bm{H}\bm{H}^H)$ the second matrix is the conjugate transpose of the first, leading to faster calculation
		\end{tablenotes}		
	\end{table}

    \begin{table}[!t]
		\centering
		\footnotesize
        \captionsetup{justification=centering}
		\caption{Matrix multiplication (MM) throughput in MM/s and \gls{gflop} at a frequency of 800~\si{\mega\hertz} on the systolic array for complex bfloat16 matrices}
		\noindent\begin{tabular}{lcc}
			\toprule
			\textbf{Matrix multiplication} & \textbf{Throughput} & \textbf{Throughput}\\ 
			\textbf{order} & \textbf{(MM/s)} & \textbf{(\gls{gflop})} \\    
			\midrule
			$(\bm{A}_{16 \times 16})$~*~$(\bm{B}_{16\times 16})$ & 10.9M & 86\\
            $(\bm{A}_{32 \times 32})$~*~$(\bm{B}_{32\times 32})$ & 2M & 129\\
            $(\bm{A}_{64 \times 64})$~*~$(\bm{B}_{64\times 64})$ & 303k & 157\\    
            $(\bm{A}_{128 \times 128})$~*~$(\bm{B}_{128\times 128})$ & 42k & 175\\  
            $(\bm{A}_{256 \times 256})$~*~$(\bm{B}_{256\times 256})$ & 5.5k & 184\\
			\bottomrule
		\end{tabular}
		\label{tab:gemms_per_second}	
	\end{table}
 
    \begin{table*}[t!]
    	\centering
    	\footnotesize
        \captionsetup{justification=centering}
    	\captionsetup{width=17cm}
    	\caption{\gls{gemm} Memory access profile for the vector core and the systolic array}
    	\noindent\begin{tabular}{l@{\hskip 0.1in}l@{\hskip 0.3in}c@{\hskip 0.3in}c@{\hskip 0.3in}c@{\hskip 0.3in}c@{\hskip 0.3in}c@{\hskip 0.3in}c}
    		\toprule
            \toprule
            
    		& \multicolumn{7}{c}{Different matrix dimensions}\\
      
    		& & $(\textbf{16} \times \textbf{16})$ & $(\textbf{32} \times \textbf{32})$ & $(\textbf{64} \times \textbf{64})$ & $(\textbf{128} \times \textbf{128})$ & $(\textbf{256} \times \textbf{256})$ & $(\textbf{16} \times \textbf{128})$\\
      
    		& & * & * & * & * & * & * \\  
      
    		& & $(\textbf{16} \times \textbf{16})$ & $(\textbf{32} \times \textbf{32})$ & $(\textbf{64} \times \textbf{64})$ & $(\textbf{128} \times \textbf{128})$ & $(\textbf{256} \times \textbf{256})$ & $(\textbf{128} \times \textbf{16})$\\    
            
    		\midrule
            \multirow{3}{*}{\rotatebox[origin=c]{90}{Systolic}}
            
    		& \#Memory read cycles & 32 & 256 & 2048 & 16.4k & 131k & 256 \\
    		& \#Memory write cycles & 16 & 64 & 256 & 1024 & 4096 & 16 \\     
    		& \#Memory cycles & 48 & 320 & 2304 & 17.4k & 135k & 272 \\       
            & \#Memory cycles / \#Execution cycles & 66\% & 80\% & 88\% & 92\% & 94\% & 89\% \\
            & & & & & & & \\
      
    		\midrule
    		\multirow{2}{*}{\rotatebox[origin=c]{90}{Vector}} 	
    		& \#Memory read cycles & 272 & 2176 & 17.4k & 139k & 1114.1k & 2176 \\
            & \#Memory write cycles & 16 & 64 & 256 & 1024 & 4096 & 16\\
            & \#Memory cycles & 288 & 2240 & 17.7k & 140k & 1118k& 2192\\
            & \#Memory cycles / \#Execution cycles & 30\% & 32\% & 33\% & 33\% & 33\% & 33\% \\
            
            \bottomrule
    		\bottomrule
    	\end{tabular}
    	\label{tab:memory_access_profile}
    \end{table*}
    
    Previously, we had designed an 8 $\times$ 8 systolic array that was embedded directly inside the extended processor pipeline with 16 stages \cite{9801536}. Although this had advantages, such as giving visibility of the internals of the systolic array to the compiler, it also came with drawbacks of much higher design complexity and lower efficiency as compared to a full-on accelerator. Consequently, for this work the systolic array has migrated out of the processor pipeline, and now acts as a fully fleshed-out, tightly-coupled accelerator. This means that the processor endorses a hands-off approach toward the systolic array, which allows the array to have full autonomy over the stewardship of the data flow.

    Fig. \ref{fig:systolic_array} depicts the block diagram of the systolic array, which is composed of three main blocks: the outer buffer registers (labeled with \mcircled{1}{black!47!green}{black!47!green}{white}{0.7}), the inner buffer registers along with their associated \gls{mac} units (marked with \mcircled{2}{red}{red}{white}{0.7}), and the systolic control unit (tagged with \mcircled{3}{blue}{blue}{white}{0.7}). 

    The outer buffer registers are loaded from the parallel vector memory, with the registers at the bottom left of the Fig.  \ref{fig:systolic_array} receiving the left-hand matrix's vectors row-wise, and the registers at the top right accepting the right-hand matrix's vectors column-wise. There is an option here to conjugate the vectors, as needed for the Gramian matrix multiplication (not shown in the figure for brevity). The register contents will then shift as ordained by the systolic control's scheduler, and the shifted values are fed into the inner buffer registers. There lies the bulk of the systolic array's arithmetic muscle with the \gls{mac} units co-located with their inner register neighbors. As was the case for the outer registers, the shifting is also controlled by the scheduler. 
    
    The systolic control is configured by custom instructions, which set up the input and output matrix addresses and dimensions, and the multiplication type (either normal \gls{gemm} or Gramian mode). To keep tabs on the matrix multiplication process, the scheduler mainly employs two counters and two bit shifters. These units, in conjunction with the information provided by the custom instructions, conspire to determine which vectors are to be put in the outer registers, which registers need to be in shifting mode, and when the blocks need to be written back to the parallel vector memory.
    
    It takes some time for the array to warm up, but once the 256 available \glspl{pe} are filled and operating concurrently, the array can proceed at full throttle, providing an immense throughput improvement over the regular vector core. Moreover, the innate local buffering within the systolic structure obviates the need for global memory accesses, and also helps to contain and simplify the internal wire routing to local connections.
    
    Table \ref{tab:systolic_savings} tabulates the performance improvements gained by the addition of the systolic array accelerator for different \gls{gemm} cases. Depending on the particular input matrix sizes, the systolic array yields an uplift of 6.3$\times$ to 23.5$\times$ in terms of the number of cycles. Table \ref{tab:gemms_per_second} quantifies the throughput capability of the system in terms of the number of matrix multiplications per second that can be performed on the systolic array for an assortment of matrix sizes.
    
    We can formulate the required number of cycles in a more formal fashion. For a \gls{gemm} of the order $({M \times N})$ times $({N \times P})$ the total number of cycles for the systolic array to finish the job is approximately calculated according to \eqref{eq:systolic_formula}
    
    \begin{eqnarray}
		\label{eq:systolic_formula}
        \frac{M}{16} \cdot \frac{P}{16} \cdot (2N \: reads + 16 \: writes + 16).
	\end{eqnarray}
    
    From \eqref{eq:systolic_formula} we can derive the numbers in Table \ref{tab:memory_access_profile}. The table shows a head-to-head comparison for the total number of memory access cycles required (broken down into reads and writes) and the percentage to the total number of execution cycles (including memory accesses). It can be seen that the systolic array achieves much better results when it comes to overlapping computation with memory accesses with up to 94\% ratio compared to the maximum of 33\% for the vector core. 

    \begin{figure}[!t]
		\centering
		\includegraphics[trim=0cm 0cm 0cm 0cm, clip=true, width=1\columnwidth]{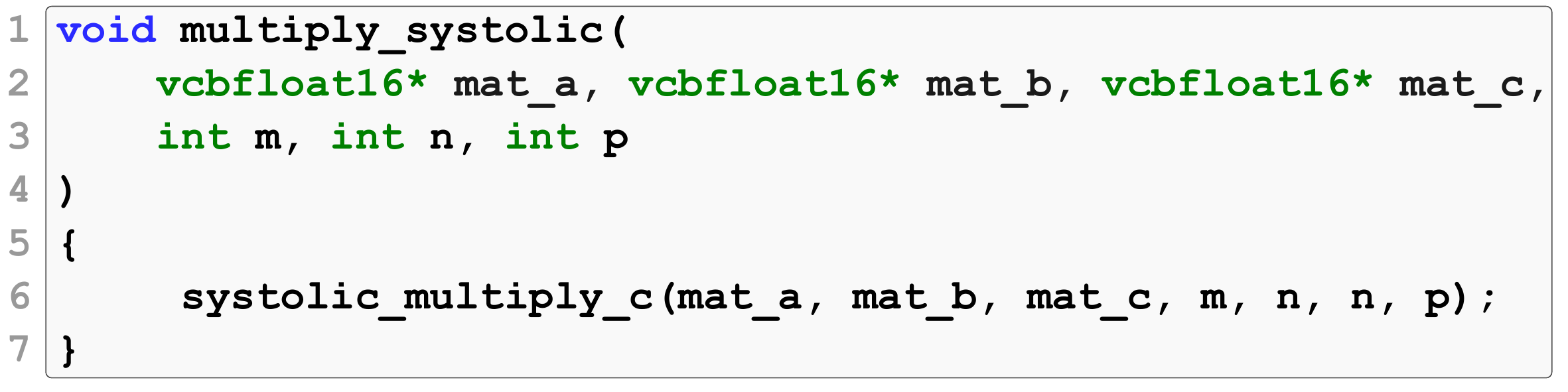}
		\vspace{\captionshift}
		\captionsetup{font={small}}
		\caption{C code with a function-like intrinsic for matrix multiplication using the systolic array accelerator}
		\label{fig:systolic_intrinsic}
	\end{figure}
 
    Now, we shift gears to see how the programmer can take advantage of the systolic array in code in an unobtrusive manner. The listing in Fig. \ref{fig:systolic_intrinsic} demonstrates the usage of the systolic array accelerator with a C language code fragment. The function \textit{multiply\_systolic} accepts input matrices \textit{mat\_a} and \textit{mat\_b} and output matrix \textit{mat\_c} which are delivered to the function as arrays of a custom complex bfloat16 vector type (\textit{vcbfloat16}). The extra arguments \textit{m}, \textit{n}, and \textit{p} determine the input matrix dimensions as $m \times n$ and $n \times p$. The \textit{systolic\_multiply\_c} at line 6 looks like a function call, but in fact is just a compiler intrinsic that helps to maintain a dialog between the programmer and the compiler. The execution  of this intrinsic acts as a trigger that sets the wheels of the systolic array in motion. What this means is that the compiler has information about this intrinsic, and whenever it comes across this name in the C code it will translate it to the relevant custom assembly instructions, which in turn initialize the parameters in systolic control as discussed earlier. This abstraction mechanism provides a much friendlier environment by insulating the programmer from all the complexities of the matrix multiplication scheduling, which are handled by the accelerator in silicon.

    \begin{figure}[!t]
		\centering
		\includegraphics[trim=0cm 0cm 0cm 0cm, clip=true, width=1\columnwidth]{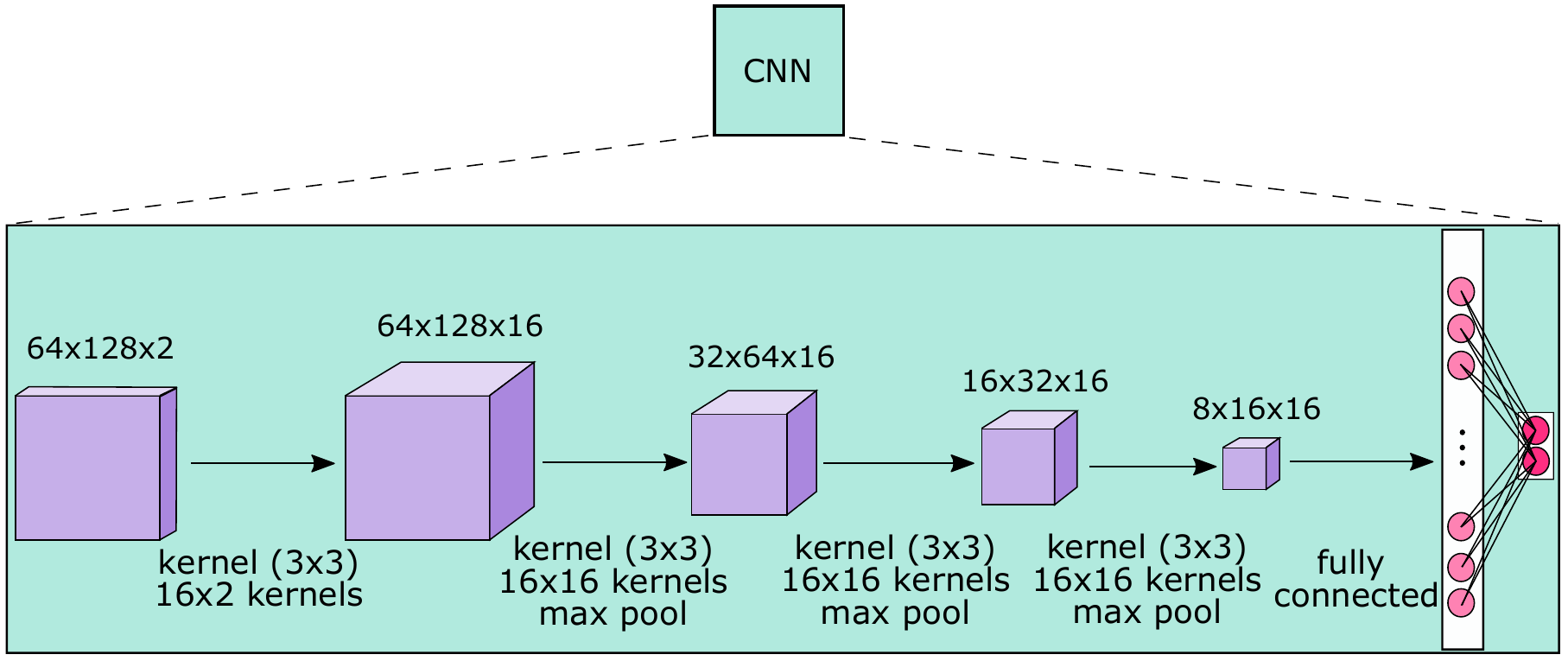}
		\vspace{\captionshift}
		\captionsetup{font={small}}
		\caption{CNN model.}
		\label{fig:cnn_model}
	\end{figure}
 
    \subsection{CNN}
    \label{subsec:cnn}
\begin{figure}[!t]
		\centering
		\includegraphics[trim=0cm 0cm 0cm 0cm, clip=true, width=1\columnwidth]{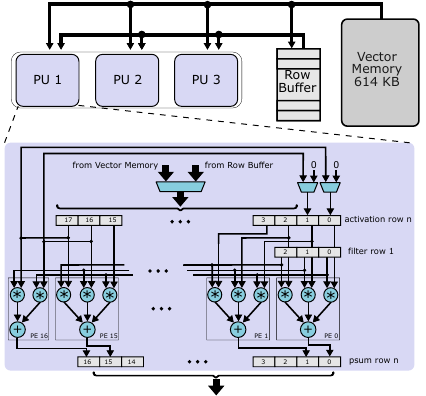}
		\vspace{\captionshift}
		\captionsetup{font={small}}
		\caption{The convolution engine of the \gls{cnn} accelerator module.}
		\label{fig:cnn}
	\end{figure}

    The \gls{cnn} module is adopted from \cite{9401528} and has been integrated in this design as the main workhorse for positioning. The \gls{cnn} model follows the structure laid out in Fig. \ref{fig:cnn_model}. The input matrix comes from a system with 128 antennas and 64 subcarriers with real and imaginary components separated, ending up with a size of 64 $\times$ 128 $\times$ 2. There are four back-to-back convolutional layers in the network culminating with a fully-connected layer. The kernel size is fixed at 3 $\times$ 3. Fig. \ref{fig:cnn} replicates the convolutional engine part of the \gls{cnn} module which employs a row-stationary dataflow model \cite{8114708}. The fixed-point implementation of the original design has been preserved, but the distinguishing factor here is that the accelerator now owns the vector memory attached to it. The coefficients used during the different stages of the \gls{cnn} processing hold court in this memory and are grabbed according to the whims of the scheduler inside the \gls{cnn} accelerator. The memory also acts as the staging area for the temporary values while working on different layers.

     \begin{figure}[!t]
		\centering
		\includegraphics[trim=0cm 0cm 0cm 0cm, clip=true, width=1\columnwidth]{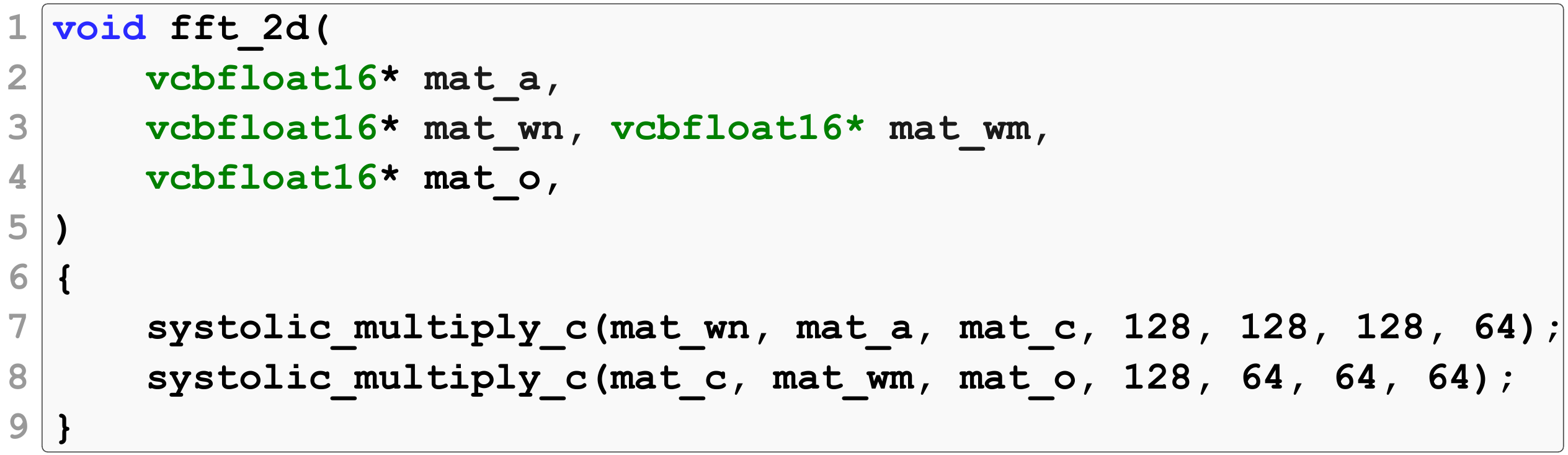}
		\vspace{\captionshift}
		\captionsetup{font={small}}
		\caption{C code for the 2D FFT.}
		\label{fig:fft_code}
	\end{figure}

    Before the \gls{cnn} module can start operating, the input matrix needs to undergo pre-processing using 2D \gls{fft}, which is carried out as two matrix multiplications by exploiting the available systolic array. So, the 2D \gls{fft} is implemented by applying 1D \gls{fft} along two dimensions, more specifically, once on the rows and once on the columns. For this we need to calculate the $N \times N$ and $M \times M$ \gls{dft} matrices $\bm{W_\text{N}}$ and $\bm{W_\text{M}}$ using

	\begin{align}
		\bm{W}_{kn} = e^{-j \frac{2\pi kn}{N}}, k,n=0,1,...,N-1,
    \end{align}
    \begin{align}
        \bm{W}_{kn} = e^{-j \frac{2\pi kn}{M}}, k,n=0,1,...,M-1.
	\end{align}    

    Armed with this knowledge, the 2D \gls{fft} of the $N \times M$ input matrix $\bm{A}$ can be obtained from pre- and post-multiplying $\bm{A}$ with the \gls{dft} matrices according to

    \begin{align}
		\bm{F} = \bm{W}_{N} \bm{A} \bm{W}^{H}_{M}.
	\end{align}

      \begin{table}[!t]
		\centering
		\footnotesize
        \captionsetup{justification=centering}
		\caption{Selected instructions from the \gls{isa}.}
		\noindent\begin{tabular}{ll}
			\toprule
			\textbf{Instruction} & \textbf{Description}\\  
			\midrule
            ldv v0, (va1++) [mode0] & Load from vector memory\\
            & with mode0 (shuffled row)\\
            stv v1, (va4) [mode1] [mask vs1] & Store to vector memory with mode1 \\
            &(shuffled column) using a mask\\
			addv v1, v2, v3 & SIMD vector addition\\
            subv v1, v2, v3 & SIMD vector subtraction\\
            mulv v1, v2, v3 & SIMD vector multiplication\\
            vmac v1, v3, v2[vs0] & Multiply and accumulate \\
            & with vector indexing\\
            vdot vs0, v2, conj(v2) & Calculate the dot product \\
            & (conjugate the2nd operand)\\
            inv.sqrt vs1, vs0 & Calculate inverse square root\\
            idxvm v0, vs0, x1 & Modify vector register at index\\
            idxv x5, v0, x4 & Read vector element at index location\\
            sys.mul (x11), (x10) & Systolic multiply with\\
            & matrix addresses provided\\
            sys.sz 1, 4, 4, 1 & Systolic matrix sizes (multiples of 16)\\
            sys.des (x11) & Systolic multiply destination address\\
			\bottomrule
		\end{tabular}
		\label{tab:instructions}	
	\end{table}
 
    Fig. \ref{fig:fft_code} lists the C code function that realizes the above method using two left and right matrix multiplications utilizing the systolic array.
 
    \begin{table*}[!h]   
        \centering
        \footnotesize
        \captionsetup{justification=centering}
        \caption{Throughput, area efficiency and energy efficiency for the developed \gls{asip}, for system dimensions of 64$\times$16, 128$\times$16 and 256$\times$16, utilizing the \gls{zf} algorithm and \gls{ofdm}}
        \centering
        \noindent\begin{tabular}{l@{\hskip 1in}ccc@{\hskip .3in}|ccc@{\hskip .3in}|ccc}
            \toprule
            System dimension & \multicolumn{3}{c|}{64$\times$16} & \multicolumn{3}{c|}{128$\times$16} & \multicolumn{3}{c}{256$\times$16}\\     
            \midrule
            {\gls{Ue} speed [\si{\kilo\meter\per\hour}]}
            & 5 & 50 & 100 & 5 & 50 & 100 & 5 & 50 & 100\\ 
            \midrule
            \midrule
            \textbf{System Parameters}& \multicolumn{6}{c}{}\\
            \midrule
            Coherence bandwidth $n_\text{b}$ [\#subcarriers]	& 16 & 16 & 16 & 16 & 16 & 16 & 16 & 16 & 16\\
            Coherence time [\si{\milli\second}]			& 7.2 & 0.73 & 0.36 & 7.2 & 0.73 & 0.36 & 7.2 & 0.73 & 0.36\\
            Coherence time $n_\text{t}$ [\#\gls{ofdm} symbols]		& 100 & 10		& 5 & 100 & 10		& 5 & 100 & 10		& 5\\
            \midrule
            \midrule
            \textbf{ASIP Performance}& \multicolumn{6}{c}{}\\
            \midrule
            Clock cycles to detect $n_\text{b}n_\text{t}$		& 26.5k & 4.4k & 3.2k & 39.7k & 5.8k & 3.9k & 66.3k & 8.6k & 5.4k \\
            Calculation time to detect $n_\text{b}n_\text{t}$ [\si{\micro\second}]   & 33.1 & 5.5 & 4.0 & 49.6 & 7.2 & 4.9 & 82.8 & 10.7 & 6.7\\ 
            \midrule
            Symbol detection throughput$^a$ [\si{\giga b\per\second}]	& 4.6 & 2.8 & 1.9 & 3.1 & 2.1 & 1.5 & 1.9 & 1.4 & 1.1\\
            Area efficiency [\si{\giga b\per s\per mm^2}] & 1.8 & 1.1 & 0.76 & 1.2 & 0.85 & 0.63 & 0.74 & 0.57 & 0.46\\
            Energy efficiency$^b$ [\si{\pico\joule\per b}] & 193 & 324 & 469 & 291 & 425 & 575 & 485 & 628 & 787\\
            \bottomrule
            \bottomrule
        \end{tabular}
        \label{tab:asip_throughput}
        \begin{tablenotes}
        \setlength\labelsep{0pt}
            \item ~~~~~~~~$^a$ Throughput = Coherence Bandwidth $\times$ \#\gls{ofdm} Symbols $\times$ \#Users (16) $\times$ \#Modulation Bits (6) / Calculation time to detect $n_\text{b}n_\text{t}$ 
            \item ~~~~~~~~$^b$ Based on estimated power consumption using switching activity results obtained from gate-level synthesized netlist simulation
        \end{tablenotes}
    \end{table*}
    
    \begin{table}[!t]
        \centering
        \footnotesize
        \captionsetup{justification=centering}
        \caption{Synthesis results with area breakdown for the \gls{asip}, vector memories and the accelerator modules}
        \noindent\begin{tabular}{lc}
            \toprule
            \textbf{Block} & \textbf{Area [\si{\square {mm}}]}\\
            \midrule
            
            \gls{cnn} vector memory (614 kB) & 0.77   \\
            Parallel vector memory (512 kB) & 0.76   \\
            Systolic array& 0.73 \\
            \gls{cnn} engine & 0.08 \\
            Vector \gls{alu} & 0.06 \\
            Instruction memory (2 kB) & 0.05  \\
            Scalar memory (2 kB) & 0.05  \\
            Miscellaneous & 0.01 \\
            \midrule
            \midrule
            Overall & 2.51\\
            \bottomrule
        \end{tabular}
        \label{tab:synthesis_area}
    \end{table}
    
    The rationale for this transformation is to bring the data into the angular-delay domain, in order to represent the channel snapshots with a sparse structure, as the \gls{cnn} machines are more efficient when the input features are sparsely distributed \cite{8292280}. Once ready, the complex matrix will be broken in two planes (that is, real and imaginary) by the memory controller and the result is copied from the parallel vector memory to the \gls{cnn} vector memory. Finally, with the real and imaginary matrices residing in the \gls{cnn} vector memory, the \gls{asip} programmatically configures the accelerator by setting up configuration registers that determine the input size and layer shapes. From this point on, and by the processor's green light, the \gls{cnn} module assumes full autonomy to carry out the positioning.
 
    \begin{table*}[!t]
        \centering
        \footnotesize
        \captionsetup{font={small}, justification=centering}
        \caption{Comparison with state-of-the-art designs. Area efficiency numbers are scaled down to a 28nm reference.\\Area efficiency\textsubscript{22nm} = (Throughput / Area) $\times$ $(22 / 28)^3$}
        \renewcommand{\arraystretch}{1.1}
        \noindent \begin{tabular}{l|ccccccc}
            \toprule
            & \textbf{This work} & Prabhu \cite{9492455} & Peng \cite{8911207} & Prabhu \cite{7870260} & Wei \cite{8310265} & Castañeda\cite{9911311} & Tan  \cite{8759904}\\
            \midrule
            Process [nm] & \textbf{22} & 28 & 28 & 28 & 28 & 22 & N/A\\
            Implementation & \textbf{Synthesis} & Tape-out & Tape-out & Tape-out & Tape-out & Tape-out & FPGA\\
            Architecture type & \textbf{\gls{asip}} & \gls{asip} & Reconfigurable & \gls{asic} & \gls{asic} & \gls{asic} & FPGA\\
            Programmability & \textbf{C-programmable} & C-programmable & N/A & N/A & N/A & C-programmable & N/A \\	
            Positioning support & \textbf{Yes} & No & No & No & No & No & No\\            
            \gls{mimo} dimension(s) & \textbf{Programmable} & 128$\times$8 & 128$\times$8 & 128$\times$8 & 128$\times$16 & Programmable & 128$\times$8\\
            Algorithm & \textbf{Linear} & \gls{zf}/\gls{mmse} & MMSE \cite{8051120} & \gls{zf}/\gls{mmse} & EPD$^a$ & Varied & Near-MMSE\\
            & \textbf{(e.g.} \textbf{\gls{zf}/\gls{mmse})} &  &  & &  &  & \\
            Datatype & \textbf{Floating-point +} & Fixed-point & Fixed-point & Fixed-point & Fixed-point & Floating-point & Fixed-point\\    
            & \textbf{fixed-point} &  &  &  &  &  & \\      \midrule     
            Frequency [MHz] & \textbf{800} & 290 & 800 & 300 & 569 & 293 & 210\\
            Power [mW] & \textbf{900} & 180 & 528 & 18 & 127 & 97 & N/A\\
            Area [\si{\square {mm}}] & \textbf{2.55}$^b$ & 1.1 & 4.8 & 1.1 & 2 & 0.42 & N/A\\ 
            Positioning rate & \textbf{\textasciitilde390} & N/A & N/A & N/A & N/A & N/A & N/A\\              
            Detection/Precoding rate & \textbf{2115}$^c$ & 169 & 1540 & 300 & 1800 & 240 & 31\\            
            ~[\si{\mega b\per s}] &  &  &  &  &  &  & \\
            \midrule
            Area efficiency & \textbf{410}$^c$ & 154 & 321 & 272 & 900 & 277 & N/A\\            
            ~[\si{\mega b\per\second\per\mm}$^2$] &  &  &  &  &  &  & \\
            Energy efficiency [\si{\pico\joule\per b}] & \textbf{425}$^c$ & 543 & 343 & 60 & 70 & 404 & N/A\\
            \bottomrule
        \end{tabular}
        \label{tab:comparison_table}
        \begin{tablenotes}
            \setlength\labelsep{0pt}
                \item $^a$ Expectation Propagation Detection, which has around 3-4 dB gain in more correlated channels compared to ZF. 
                \item $^b$ Includes memories\\
                \item $^c$ For a 128$\times$16 massive \gls{mimo} system with a coherence time of 10 \gls{ofdm} symbols and 64-\gls{qam}\\
        \end{tablenotes}			
    \end{table*}
 
    \subsection{Instructions}
    In the preceding sections we visited the challenges that crop up in communications and positioning and then unpicked the computational demands that are put on the processing system, while providing a basket of kernel operations that ought to be implemented in such systems. We decided to go with an \gls{asip} as the main conduit for providing flexibility and married it to custom accelerators as the scaffolding around it. Now, let us take a brief look at the instructions that exemplify this customization to drive the point home. Table \ref{tab:instructions} puts together a sample list of instructions implemented in the \gls{asip} that ties directly to the discussions up to this point. The next section devotes itself to evaluating the design and implementation results.
 
	\section{Evaluation and Implementation Results}
	\label{sec:evaluation_sec}
	\subsection{Algorithm Run-time Analysis and System Performance}
    The processor is developed with the \gls{asip} Designer \cite{asip_designer} tool from Synopsys, and is programmable in the C programming language. The tool keeps the compiler in the loop during design time and furnishes handy hardware-expressive features. Furthermore, it comes with a cycle-accurate simulator which is utilized to obtain the following evaluation results. The benchmarks are for \gls{ul} detection for massive \gls{mimo} systems utilizing the \gls{zf} algorithm with different dimensions.
    
    For positioning, the \gls{cnn} input tensor size is $K \times M \times 2$, in which 2 represents the real and imaginary components of the complex input\footnote{For this work we consider real-valued \gls{cnn} processing only, thus real and imaginary parts are fed to the network as two separate input channels.}. For evaluation purposes, we assume the number of antennas to be $M=128$, and the number of subcarriers to be $K=64$, resulting in an input to the CNN of dimensions $64 \times 128 \times 2$. The network is composed of four convolutional layers and a fully connected layer at the end. The network utilizes the \gls{relu} to introduce non-linearity. There are 16 filters per convolutional layer, with 2 kernels per filter for the first layer (to match real and imaginary channels), and 16 kernels per filter afterwards, with a kernel size of $3 \times 3$ applied across all the layers \cite{9401528}.

    The overall system performance in terms of detection time, throughput, area efficiency and energy efficiency are quantified in Table \ref{tab:asip_throughput}, assuming a coherence bandwidth of 16 subcarriers and a 64-\gls{qam} constellation scheme. The results are given for the \gls{ue} speeds of 5, 50, and 100 \si{\kilo\meter\per\hour}. A 128 $\times$ 16 massive \gls{mimo} setup achieves a maximum throughput of 2.1 Gb/s for the 50 \si{\kilo\meter\per\hour} case. Notably, the system can keep up with frequent \gls{csi} estimations (e.g. every 1 ms).
    The processor takes roughly 2M cycles to perform one localization, of which a significant portion is spent on the \gls{cnn} calculation. This means a user terminal's position can be determined every 2.5 milliseconds (for an 800 MHz clock), with a positioning error of 40 cm on average.
    
    \subsection{Synthesis Results}
    The design has been synthesized using the \SI{22}{\nano\meter} \gls{fd-soi} process node, and is fully verified against the cycle-accurate simulator. A real estate breakdown of the different components in the system is provided in Table \ref{tab:synthesis_area}. The whole system takes up an area of around 2.5 $\mathrm{mm^2}$, synthesized with an operating frequency of 800 MHz. The systolic array and the two vector memories assume the bulk of the area. The rest of the modules pale in comparison and sit apart at below 0.1 $\mathrm{mm^2}$.
    
    Table \ref{tab:comparison_table} summarizes the comparison results by pitting our \gls{asip} design against the state of the art, which includes other \gls{asip}, configurable, and \gls{asic} implementations appearing in \cite{9492455, 8911207, 7870260, 8310265, 9911311}. Despite being programmable, the proposed \gls{asip} system can closely compete with the \gls{asic} designs. 
  
	\section{Conclusion}
	\label{sec:conclusion}
    Creating compute systems that meet the needs of the complex communications networks of the present and posterity is a tall order for system architects. This has compelled designers to shift thier focus towards algorithms/software performance-engineering along with specialization of computer architecture, to forge a viable path forward. In this paper we investigated the utilization of an ASIP vector processor, allied with a tightly-coupled systolic array accelerator to tackle the compute-heavy task of detection and precoding for massive \gls{mimo} systems. Moreover, a \gls{cnn} module accelerator further beefs up the processor to undertake the positioning functionality. The programmable processor achieves a post-synthesis frequency of 800 MHz, and secures a maximum detection throughput of 2.1 Gb/s in a 128$\times$16 massive \gls{mimo} system for the \gls{ue} speed of 50\si{\kilo\meter\per\hour}, with an estimated power draw of 900 mW. As for the positioning, it can churn out around 390 user positions per second, with an average distance error of 3.5 $\lambda$.
  
	\section*{Acknowledgment}
    The authors would like to thank Synopsys for providing their tool ASIP Designer.

	\bibliographystyle{IEEEtran}
    \bibliography{main.bbl}

\begin{thebibliography}{10}
\providecommand{\url}[1]{#1}
\csname url@samestyle\endcsname
\providecommand{\newblock}{\relax}
\providecommand{\bibinfo}[2]{#2}
\providecommand{\BIBentrySTDinterwordspacing}{\spaceskip=0pt\relax}
\providecommand{\BIBentryALTinterwordstretchfactor}{4}
\providecommand{\BIBentryALTinterwordspacing}{\spaceskip=\fontdimen2\font plus
\BIBentryALTinterwordstretchfactor\fontdimen3\font minus \fontdimen4\font\relax}
\providecommand{\BIBforeignlanguage}[2]{{%
\expandafter\ifx\csname l@#1\endcsname\relax
\typeout{** WARNING: IEEEtran.bst: No hyphenation pattern has been}%
\typeout{** loaded for the language `#1'. Using the pattern for}%
\typeout{** the default language instead.}%
\else
\language=\csname l@#1\endcsname
\fi
#2}}
\providecommand{\BIBdecl}{\relax}
\BIBdecl

\bibitem{10.1145/3282307}
\BIBentryALTinterwordspacing
J.~L. Hennessy and D.~A. Patterson, ``A new golden age for computer architecture,'' \emph{Commun. ACM}, vol.~62, no.~2, p. 48–60, jan 2019. [Online]. Available: \url{https://doi.org/10.1145/3282307}
\BIBentrySTDinterwordspacing

\bibitem{8416771}
L.~Van~der Perre, L.~Liu, and E.~G. Larsson, ``{Efficient DSP and Circuit Architectures for Massive MIMO: State of the Art and Future Directions},'' \emph{IEEE Transactions on Signal Processing}, vol.~66, no.~18, pp. 4717--4736, 2018.

\bibitem{9319703}
S.~Shahabuddin, A.~Mämmelä, M.~Juntti, and O.~Silvén, ``{ASIP for 5G and Beyond: Opportunities and Vision},'' \emph{IEEE Transactions on Circuits and Systems II: Express Briefs}, vol.~68, no.~3, pp. 851--857, 2021.

\bibitem{7446051}
T.~Nowatzki, V.~Gangadhan, K.~Sankaralingam, and G.~Wright, ``Pushing the limits of accelerator efficiency while retaining programmability,'' in \emph{2016 IEEE International Symposium on High Performance Computer Architecture (HPCA)}, 2016, pp. 27--39.

\bibitem{9492455}
H.~Prabhu, L.~Liu, F.~Sheikh, and O.~Edfors, ``{A 1070 pJ/b 169 Mb/s Quad-core Digital Baseband SoC for Distributed and Cooperative Massive MIMO in 28 nm FD-SOI},'' in \emph{2021 Symposium on VLSI Circuits}, 2021, pp. 1--2.

\bibitem{8911207}
G.~Peng, L.~Liu, S.~Zhou, S.~Yin, and S.~Wei, ``{A 2.92-Gb/s/W and 0.43-Gb/s/MG Flexible and Scalable CGRA-Based Baseband Processor for Massive MIMO Detection},'' \emph{IEEE Journal of Solid-State Circuits}, vol.~55, no.~2, pp. 505--519, 2020.

\bibitem{7870260}
H.~Prabhu, J.~N. Rodrigues, L.~Liu, and O.~Edfors, ``{3.6 A 60pJ/b 300Mb/s 128×8 Massive MIMO precoder-detector in 28nm FD-SOI},'' in \emph{2017 IEEE International Solid-State Circuits Conference (ISSCC)}, 2017, pp. 60--61.

\bibitem{8310265}
W.~Tang, H.~Prabhu, L.~Liu, V.~Öwall, and Z.~Zhang, ``{A 1.8Gb/s 70.6pJ/b 128×16 link-adaptive near-optimal massive MIMO detector in 28nm UTBB-FDSOI},'' in \emph{2018 IEEE International Solid - State Circuits Conference - (ISSCC)}, 2018, pp. 224--226.

\bibitem{9911311}
O.~Castañeda, L.~Benini, and C.~Studer, ``{A 283 pJ/b 240 Mb/s Floating-Point Baseband Accelerator for Massive MU-MIMO in 22FDX},'' in \emph{ESSCIRC 2022- IEEE 48th European Solid State Circuits Conference (ESSCIRC)}, 2022, pp. 357--360.

\bibitem{Waterman:EECS-2016-118}
\BIBentryALTinterwordspacing
A.~Waterman, Y.~Lee, D.~A. Patterson, and K.~Asanović, ``{The RISC-V Instruction Set Manual, Volume I: User-Level ISA, Version 2.1},'' EECS Department, University of California, Berkeley, Tech. Rep. UCB/EECS-2016-118, May 2016. [Online]. Available: \url{http://www2.eecs.berkeley.edu/Pubs/TechRpts/2016/EECS-2016-118.html}
\BIBentrySTDinterwordspacing

\bibitem{10049118}
E.~Cui, T.~Li, and Q.~Wei, ``{RISC-V Instruction Set Architecture Extensions: A Survey},'' \emph{IEEE Access}, vol.~11, pp. 24\,696--24\,711, 2023.

\bibitem{9801536}
M.~Attari, L.~Ferreira, L.~Liu, and S.~Malkowsky, ``{An Application Specific Vector Processor for Efficient Massive MIMO Processing},'' \emph{IEEE Transactions on Circuits and Systems I: Regular Papers}, vol.~69, no.~9, pp. 3804--3815, 2022.

\bibitem{9401528}
M.~Attari, J.~R. Sánchez, L.~Liu, and S.~Malkowsky, ``{An Application Specific Vector Processor for CNN-Based Massive MIMO Positioning},'' in \emph{2021 IEEE International Symposium on Circuits and Systems (ISCAS)}, 2021, pp. 1--5.

\bibitem{8292280}
J.~Vieira, E.~Leitinger, M.~Sarajlic, X.~Li, and F.~Tufvesson, ``Deep convolutional neural networks for massive mimo fingerprint-based positioning,'' in \emph{2017 IEEE 28th Annual International Symposium on Personal, Indoor, and Mobile Radio Communications (PIMRC)}, 2017, pp. 1--6.

\bibitem{10330061}
G.~Tian, I.~Yaman, M.~Sandra, X.~Cai, L.~Liu, and F.~Tufvesson, ``Deep-learning-based high-precision localization with massive mimo,'' \emph{IEEE Transactions on Machine Learning in Communications and Networking}, vol.~2, pp. 19--33, 2024.

\bibitem{5995137}
J.~Janhunen, T.~Pitkanen, O.~Silven, and M.~Juntti, ``{Fixed- and Floating-Point Processor Comparison for MIMO-OFDM Detector},'' \emph{IEEE Journal of Selected Topics in Signal Processing}, vol.~5, no.~8, pp. 1588--1598, 2011.

\bibitem{intel_amx}
\BIBentryALTinterwordspacing
``{Accelerate Artificial Intelligence (AI) Workloads with Intel Advanced Matrix Extensions (Intel AMX)}.'' [Online]. Available: \url{https://www.intel.com/content/www/us/en/content-details/785250/accelerate-artificial-intelligence-ai-workloads-with-intel-advanced-matrix-extensions-intel-amx.html}
\BIBentrySTDinterwordspacing

\bibitem{google_bfloat16_tpu}
\BIBentryALTinterwordspacing
``{BFloat16: The secret to high performance on Cloud TPUs}.'' [Online]. Available: \url{https://cloud.google.com/blog/products/ai-machine-learning/bfloat16-the-secret-to-high-performance-on-cloud-tpus}
\BIBentrySTDinterwordspacing

\bibitem{cerebras_bfloat16}
\BIBentryALTinterwordspacing
``{To Bfloat or not to Bfloat? That is the Question!}'' [Online]. Available: \url{https://www.cerebras.net/machine-learning/to-bfloat-or-not-to-bfloat-that-is-the-question/}
\BIBentrySTDinterwordspacing

\bibitem{google_bfloat16}
\BIBentryALTinterwordspacing
``The bfloat16 numerical format.'' [Online]. Available: \url{https://cloud.google.com/tpu/docs/bfloat16}
\BIBentrySTDinterwordspacing

\bibitem{10405907}
S.~Cheng, O.~A. Topal, M.~Ozger, C.~Cavdar, O.~Edfors, and L.~Liu, ``A hardware-in-the-loop simulator for mmwave massive mimo using pynq framework,'' in \emph{2023 IEEE 66th International Midwest Symposium on Circuits and Systems (MWSCAS)}, 2023, pp. 698--702.

\bibitem{Yang:2017}
Y.~{Liu}, L.~{Liu}, and V.~{Öwall}, ``{Architecture Design of a Memory Subsystem for Massive MIMO Baseband Processing},'' \emph{IEEE Transactions on Very Large Scale Integration (VLSI) Systems}, vol.~25, no.~10, pp. 2976--2980, 2017.

\bibitem{8445578}
Y.~Liu, L.~Liu, O.~Edfors, and V.~Öwall, ``{An Area-Efficient On-Chip Memory System for Massive MIMO Using Channel Data Compression},'' \emph{IEEE Transactions on Circuits and Systems I: Regular Papers}, vol.~66, no.~1, pp. 417--427, 2019.

\bibitem{8839449}
S.~Savas, Y.~Atwa, T.~Nordström, and Z.~Ul-Abdin, ``Using harmonized parabolic synthesis to implement a single-precision floating-point square root unit,'' in \emph{2019 IEEE Computer Society Annual Symposium on VLSI (ISVLSI)}, 2019, pp. 621--626.

\bibitem{bb5afff9714941c681fc0e5270168a71}
W.~Hwu, D.~Kirk, and I.~Hajj, \emph{\BIBforeignlanguage{English (US)}{Programming Massively Parallel Processors: A Hands-on Approach, Fourth Edition}}.\hskip 1em plus 0.5em minus 0.4em\relax Elsevier, Jan. 2022, publisher Copyright: {\textcopyright} 2023 Elsevier Inc. All rights reserved.

\bibitem{kung1979systolic}
H.~T. Kung and C.~E. Leiserson, ``{Systolic arrays (for VLSI)},'' in \emph{Sparse Matrix Proceedings 1978}, vol.~1.\hskip 1em plus 0.5em minus 0.4em\relax Society for industrial and applied mathematics Philadelphia, PA, USA, 1979, pp. 256--282.

\bibitem{kung1982systolic}
H.~T. Kung, ``Why systolic architectures?'' \emph{Computer}, vol.~15, no.~1, pp. 37--46, 1982.

\bibitem{doi:10.1080/00207169808804678}
\BIBentryALTinterwordspacing
C.~R. Wan and D.~J. Evans, ``Nineteen ways of systolic matrix multiplication,'' \emph{International Journal of Computer Mathematics}, vol.~68, no. 1-2, pp. 39--69, 1998. [Online]. Available: \url{https://doi.org/10.1080/00207169808804678}
\BIBentrySTDinterwordspacing

\bibitem{8114708}
V.~{Sze}, Y.~{Chen}, T.~{Yang}, and J.~S. {Emer}, ``{Efficient Processing of Deep Neural Networks: A Tutorial and Survey},'' \emph{Proceedings of the IEEE}, vol. 105, no.~12, pp. 2295--2329, 2017.

\bibitem{8759904}
X.~Tan, J.~Jin, K.~Sun, Y.~Xu, M.~Li, Y.~Zhang, Z.~Zhang, X.~You, and C.~Zhang, ``{Enhanced Linear Iterative Detector for Massive Multiuser MIMO Uplink},'' \emph{IEEE Transactions on Circuits and Systems I: Regular Papers}, vol.~67, no.~2, pp. 540--552, 2020.

\bibitem{8051120}
G.~Peng, L.~Liu, S.~Zhou, S.~Yin, and S.~Wei, ``{A 1.58 Gbps/W 0.40 Gbps/mm2 ASIC Implementation of MMSE Detection for $128\times 8~64$ -QAM Massive MIMO in 65 nm CMOS},'' \emph{IEEE Transactions on Circuits and Systems I: Regular Papers}, vol.~65, no.~5, pp. 1717--1730, 2018.

\bibitem{asip_designer}
\BIBentryALTinterwordspacing
Synopsys, \emph{{ASIP Designer}}. [Online]. Available: \url{https://www.synopsys.com/dw/ipdir.php?ds=asip-designer}
\BIBentrySTDinterwordspacing

\end{thebibliography}

	\begin{IEEEbiography}[{\includegraphics[width=1in,height=1.25in,clip,keepaspectratio]{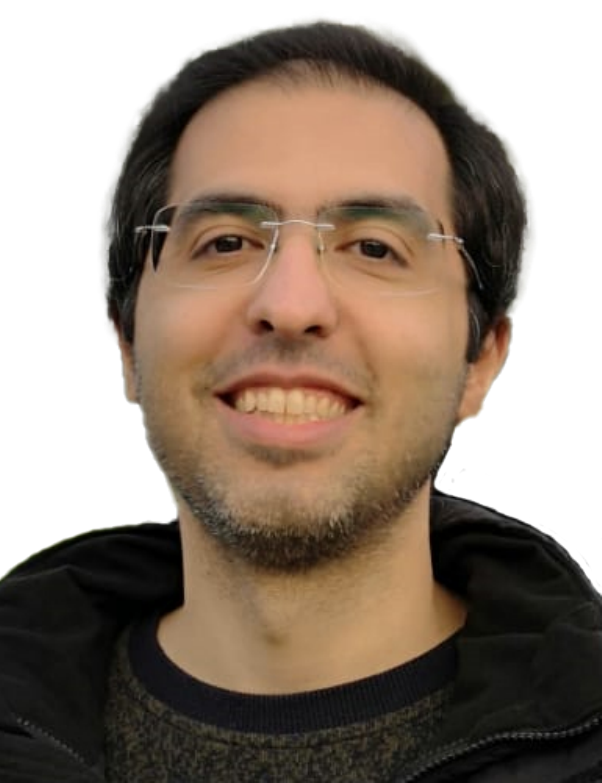}}]{Mohammad Attari}
		 (Student Member, IEEE) was born in Iran in 1983. He received his B.Eng. in Computer Hardware Engineering from Iran University of Science and Technology (IUST) in 2008, and holds an M.Sc. in Integrated Electronics System Design from Chalmers University of Technology, which he received in 2012. After his studies, he took up a position in ARM (Lund, Sweden) in the video processing division as an RTL developer. He is currently a PhD student in the Digital ASIC research group in the Electrical and Information Technology (EIT) department at Lund University. His research focuses on developing massive MIMO and next generation baseband communications processors. His interests include (but are not limited to) computer architecture, domain specific architectures, accelerator-level parallelism, application specific instruction set processors (ASIP), neural networks, and cross-level hardware-software (-algorithm) codesign and implementation.
	\end{IEEEbiography}

	\begin{IEEEbiography}[{\includegraphics[width=1in,height=1.25in,clip,keepaspectratio]{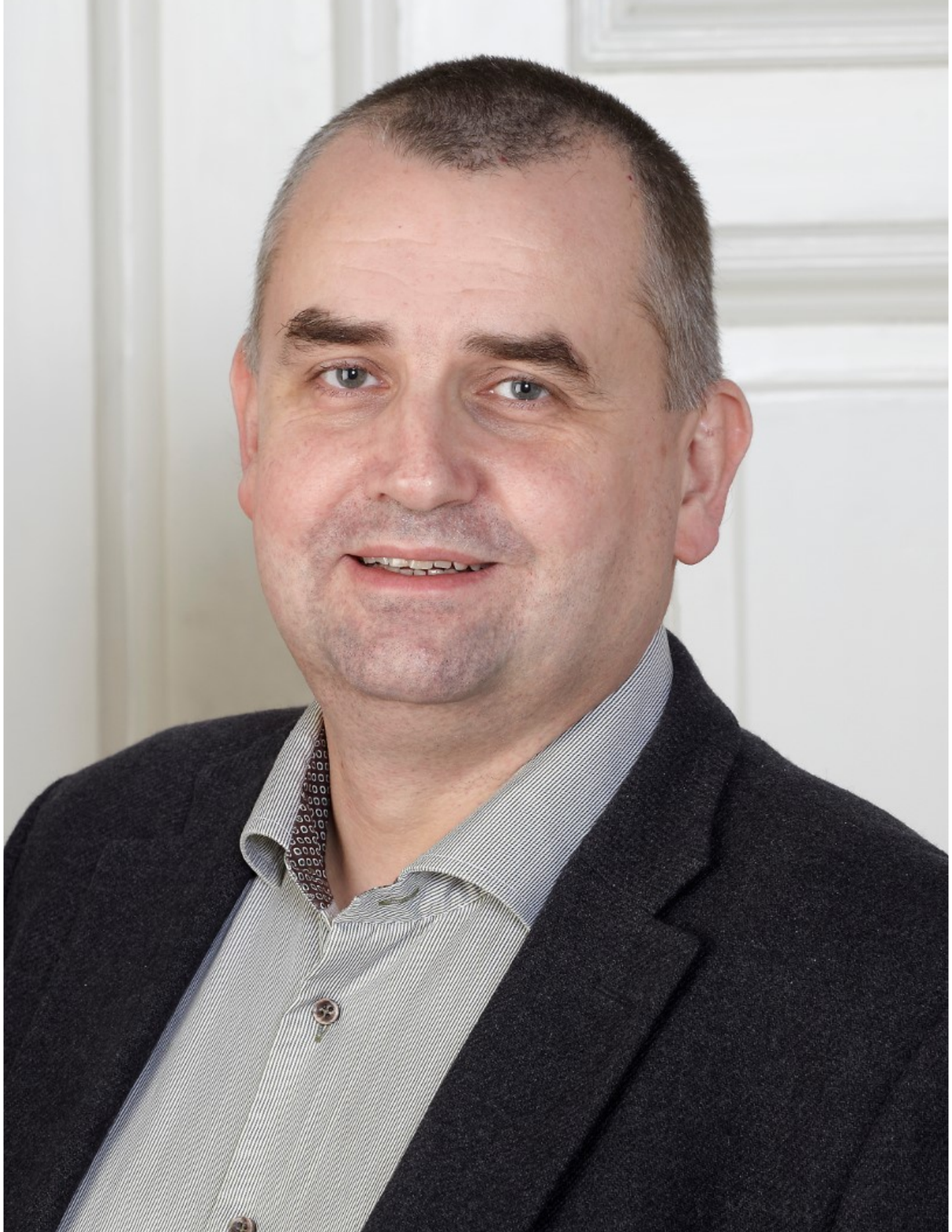}}]{Ove Edfors}
        (Senior Member, IEEE) is currently a professor in radio systems with the Department of Electrical and Information Technology, Lund University, Lund, Sweden. His current research interests include statistical signal processing and low-complexity algorithms with applications in wireless communications. In the context of massive MIMO and large intelligent surfaces, his main research interests include how realistic propagation characteristics influence system performance and baseband processing complexity.
	\end{IEEEbiography}
 
	\begin{IEEEbiography}[{\includegraphics[width=1in,height=1.25in,clip,keepaspectratio]{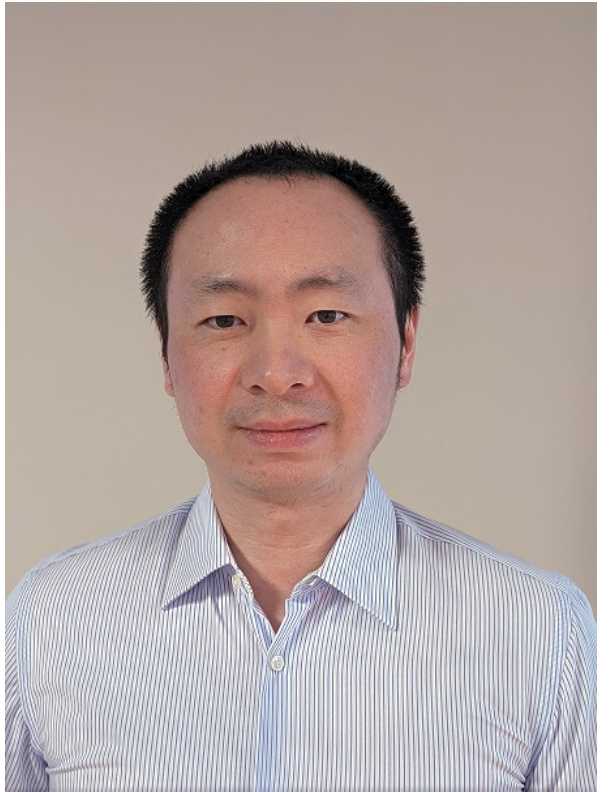}}]{Liang Liu}
        (Member, IEEE) 
 		 received the B.S. degree from the Department of Electronics Engineering, Fudan University, Shanghai, China, in 2005, and the Ph.D. degree from the Department of Microelectronics, Fudan University in 2010. In 2010, he was a Visiting Researcher with the Rensselaer Polytechnic Institute, USA. He joined the Department of Electrical and Information Technology (EIT), Lund University, Lund, Sweden, where he held a postdoctoral position in 2010. Since 2016, he has been an Associate Professor with the Department of Electrical and Information Technology (EIT), Lund University. His current research interests include wireless communication systems and digital integrated circuits design. Dr. Liu is a Board Member of the IEEE Swedish SSC/CAS Chapter. He is also a member of the Technical Committee of VLSI Systems and Applications and CAS for Communications of the IEEE Circuit and Systems Society.
	\end{IEEEbiography}

\end{document}